\documentclass{article}

\usepackage{amsmath}
\usepackage{graphicx}
\usepackage{subcaption}
\usepackage{natbib}
\usepackage{hyperref}       
\usepackage{geometry}       
\usepackage{xcolor}

\title{HALO: A High-Precision Orbit Propagation Tool for Mission Design in the Cis-Lunar Domain}
\author{
  Quentin Granier$^{1,2}$, Yang Yang$^{2,}$\thanks{Corresponding author: yang.yang16@unsw.edu.au}, and Andrew Dempster$^{1}$ \\
  \\
  \small{$^1$Australian Center for Space Engineering Research,} \\ \small{School of Electrical Engineering and Telecommunications, UNSW, Sydney, NSW 2052, Australia,} \\
  \small{$^3$School of Mechanical and Manufacturing Engineering, UNSW, Sydney, NSW 2052, Australia}
}
\date{}

\begin{document}

\maketitle

\begin{abstract}
With the recent implementation of the Artemis Accords, interest in the cis-lunar space is rapidly increasing, necessitating the development of more precise and accurate modeling tools. While general-purpose mission design tools are available, this study proposes an open-source mission design tool, HALO, for High-precision Analyser for Lunar Orbits. This work presents a comprehensive review of the modeling approaches, structural design, and algorithms employed, aiming at facilitating ease of use and adaptation for other research in the cis-lunar domain. Furthermore, accuracy studies of the propagator are provided for various orbits of interest within this domain, including low lunar orbits, elliptical frozen orbits, and 3-body problem orbits, such as Near Rectilinear Halo Orbits and Distant Retrograde Orbits.
\end{abstract}


\section{Introduction}

This study is situated within the framework of the Artemis Accords, initiated by the United States and endorsed by 43 parties globally, including Australia since its inception in 2020. These accords establish a framework for international collaboration in the civil exploration and peaceful utilization of extraterrestrial bodies and objects. The Accords were developed in conjunction with US's National Aeronautics and Space Administration's (NASA) 2017 Artemis program, which involves a budget of US\$100 billion and aims to return humanity to the Moon while advancing cis-lunar space exploration. Mastery of the cis-lunar domain is intended to establish a permanent human presence on the Moon and facilitate access to deep space, thereby advancing the future of global human exploration to Mars and beyond.

Recent years have witnessed the development of various projects in this domain, including the Gateway mission \citep{Gateway}, the Artemis missions \citep{ArtemisI}, and the Japan Aerospace Exploration Agency (JAXA) Lunar Navigation Satellite System (LNSS) \citep{ELFO}. These projects involve different orbits of interest within the cis-lunar domain. The Gateway mission intends to establish a lunar space station in a Southern Near Rectilinear Halo Orbit (NRHO) around the L2 Lagrange point, an orbit characterized by the gravitational interaction between the Earth and the Moon. The Capstone mission, launched in June 2022, was tasked with exploring this future lunar space station orbit \citep{Capstone}. Additionally, Artemis I, launched in November 2022, investigated the Distant Retrograde Orbit (DRO), another type of orbit governed by three-body dynamics. Meanwhile, JAXA's LNSS is developing an efficient navigation and communication system around the Moon, utilizing Elliptical Lunar Frozen Orbits (ELFO). The lunar gravitational field (LGF), which is complex due to geological history and mass concentrations (mascons) \citep{Mascons}, significantly affects orbits close to the Moon, often causing them to decay. Some orbits show periodic changes and are then called ``frozen" orbits like the ELFO. Given the complexity of the LGF, low lunar orbit (LLO) also warrants detailed study. This study will consider the four aforementioned orbits as reference points within the cis-lunar domain.

\begin{figure}[!ht]
    \centering
    \includegraphics[width=1\linewidth]{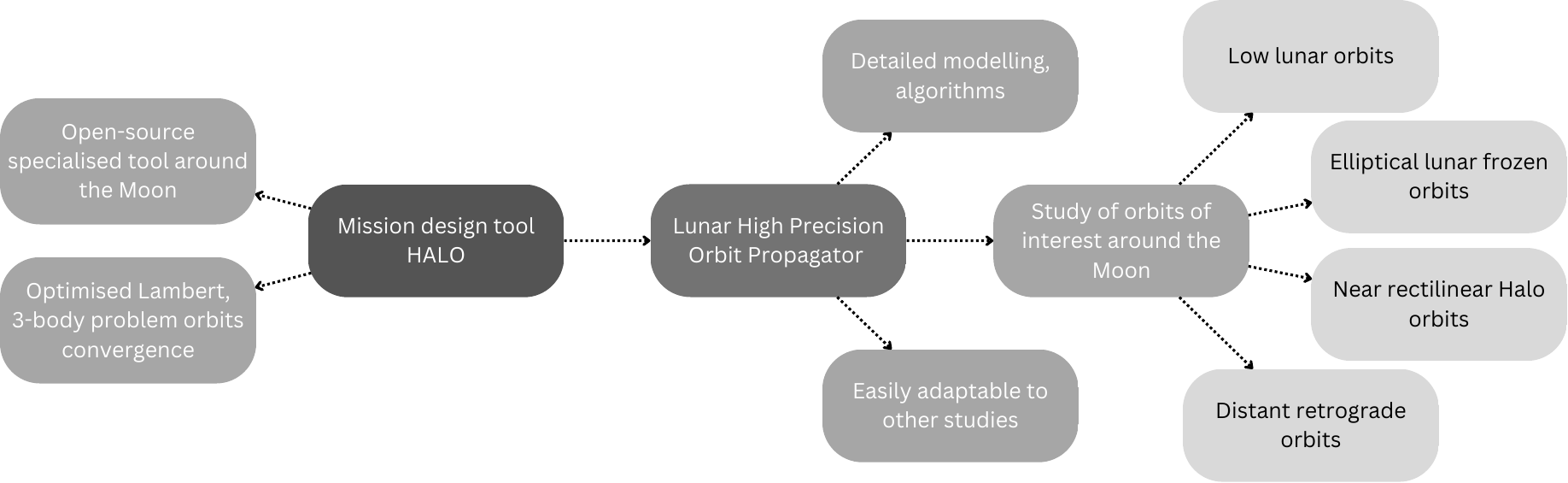}
    \caption{Graphical table of contents}
    \label{fig:GToC}
\end{figure}

As previously discussed, the cis-lunar domain is governed by multiple complex forces, and the increasing interest in this area necessitates the development of lunar high-precision orbit propagators (LHPOP). Existing tools such as the Systems Tool Kit (STK from Ansys) and the General Mission Analysis Tool \citep{GMAT} (GMAT from NASA) are general-purpose software; although GMAT is open-source, modifications and precise modeling within it are not easily accessible. This study addresses the need for a specialized LHPOP tailored to lunar orbits, offering a comprehensive review of the modeling techniques, sensitivity analysis on cis-lunar domain orbits, and integration into a mission design tool with detailed algorithms. The tool is named as HALO for High-precision Analyser for Lunar Orbits and is accessible in open source at https://github.com/Quent2G/High-precision-Analyser-of-Lunar-Orbits. In addition to orbit propagation, HALO enables the resolution of mission design problems, such as the optimization of Lambert transfers or the convergence of three-body problem (TBP) orbits from simplified circular restricted models to highly accurate ephemeris models. Figure \ref{fig:GToC} summarizes the key considerations of the study with a graphical table of content.

The main contributions by this work are summarized as follows:
\begin{itemize}
    \item HALO is a specialized tool designed for precise lunar orbit modeling and mission design, offering detailed algorithms and flexibility through open-source access, eliminating the limitations of using GMAT or STK.
    \item HALO enables detailed analysis of LLO, ELFO, NRHO, and DRO, providing insights into orbit dynamics and sensitivities for mission planning.
    \item HALO's lunar orbit propagator is validated with spacecraft ephemerides, ensuring accurate and reliable orbit predictions for cis-lunar missions.
\end{itemize}

The remainder of this paper is organized as follows: Section \ref{TaCS} defines the temporal considerations and coordinate systems utilized. Section \ref{Modelling} discusses the models employed and the enhancements implemented. Section \ref{Assess} presents the accuracy and performance outcomes for the aforementioned orbits, and Section \ref{MD} introduces the mission design tool and the algorithms used, followed by concluding remarks and acknowledgements.

\section{Time and coordinate systems} \label{TaCS}
Time and coordinate system considerations are crucial for any space-related application. The concept of time on Earth varies by time zone, depends on the Earth's rotational velocity, and is not continuous, with the second as the smallest unit. When factoring in general relativistic effects, it becomes necessary to establish a precise and universally applicable time reference by fixing a specific location and epoch and considering a continuous progression of time independent of physical influences. Similarly, while different orbits or forces may be best defined within specific reference frames, a common inertial frame is required to integrate the equations of motion. This section addresses the time and frame considerations to establish a solid foundation for the subsequent analyses. More information on the implementation can be found in the respective Section \ref{sec:TimeCoord}.

\subsection{Time}
When processing orbital calculation, one has to be more precise than the second and the discontinuous calendar time is not sufficient. To this extent, the Barycentric Dynamic Time (TDB) also referred to as the Ephemeris Time (ET) can be used. The TDB would be the time that is kept by an atomic clock at the solar system barycenter. It is defined as the independent variable in the differential equation of motion of all solar system planets \citep{SatOrbMod}. This number corresponds to the number of ephemeris seconds past J2000, so the seconds (and part seconds) elapsed if we were at the solar system barycenter. Specifically, J2000 corresponds to the epoch of the 1st January 2000 at noon (12pm). It is important to note that this epoch is referenced in TDB and not in UTC (Coordinated Universal Time), being the Earth time at Greenwich globally used in everyday life after consideration of the time zones. Those times are different (but really close) because of the relativistic effect and the leap seconds. The ``leap seconds" are seconds added every once in a while to the UTC time so that the UTC noon matches the noon solar alignment.

\subsection{Frames}
\subsubsection{J2000 for integration}
Considering a mission design tool, some forces or considerations can exist in different specific frames. For instance, the geopotential calculations for a specific body have to be done in a body fixed frame but the integration has to be done in an inertial frame of reference. Every force has then to be converted in the integration frame. Usually we consider the J2000 frame as the reference frame for integration, which corresponds to the Earth's mean equator and equinox at the 1 Jan 2000 12pm TDB.

\subsubsection{Body-fixed frames}
As mentioned before, in the computation of the gravity fields around a body, one has to consider the position of the spacecraft in a body-fixed frame. The propagator considers the gravity fields of two bodies, i.e., the Moon and the Earth, and the associated coefficients are defined in a specific frame. 

For the Moon, the selenocentric Principal Axes (PA) frame is used. The principal axes are defined as the axes for which the Moon's inertia tensor is diagonal. They correspond to the axes around which the body tends to rotate.

For the Earth, the geocentric World Geodetic System 84 (WGS84) frame designed for the Global Positioning System is used. Some other International Terrestrial Reference Frames can also be used, such as the ITRF93 used in SPICE\footnote{SPICE will be introduced in the implementation Section \ref{sec:TimeCoord}. It helps managing the time and frames related considerations} (Spacecraft, Planet, Instrument, C-matrix and Events), which is aligned with the WGS84 to within less than 0.1m on Earth surface \citep{WGSITRF}.

\subsubsection{Earth-Moon rotational frame for TBP orbits} \label{3BPO}
The study of the special TBP orbits like the NRHOs or the DROs leads to the use of the Earth-Moon rotational frame. Indeed, those TBP orbits can be computed in a simplified analytic model called the Circular Restricted 3 Body Problem (CR3BP). This model considers a circular motion of the Moon around the Earth and the equation are in the rotating frame following the movement of the Moon. This frame is defined by its center on the Earth-Moon barycenter, the x-axis pointing towards the Moon, the y-axis in the direction of the Moon's velocity and the z-axis orthogonal to the orbital plane.

\begin{figure}[!ht]
    \centering
    \includegraphics[width=0.5\linewidth]{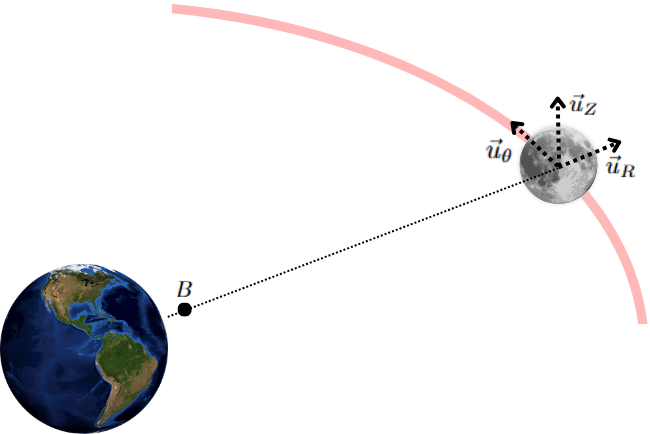}
    \caption{Definition of the rotational frame from the inertial one}
    \label{fig:Rot}
\end{figure}

In an inertial Moon centered frame like the integration frame, those axes can be represented respectively by $u_{\text{R}}$, $u_\theta$ and $u_{\text{Z}}$ shown on Figure \ref{fig:Rot}. Two main cases can lead to a transition between the inertial frame and the rotational one. First of all, after the computation of a CR3BP solution, we would like to simulate the same initial state with our high precision propagator and we need to convert this initial state from a circular restricted model in the rotational frame to a precise ephemeris model in the inertial frame. The second case would be when we propagate our initial condition, all the computations are done in the integration frame, an inertial frame. We may then also want to visualise the resulting propagation in the rotational frame. The first case requires to fit the simplified model on an ephemeris model followed by the conversion into the inertial frame. The second case only requires to do the inverse conversion, without the fitting part. The first case will be considered first, from which the second case will be derived.

First, since the CR3BP model is circular restricted, some assumptions must be made in order to fit it to an ephemeris model. With the Moon as the primary body, we can define that the Moon is at the same place in both models. Consequently, the Earth-Moon distance is fixed to an average lunar unit (LU) of $R_{\text{LU}} = 389703$ km. Therefore, the Earth cannot stay at the same position in both models. We then define the CR3BP Earth to be in the same direction as the ephemeris Earth but at a distance of $R_{\text{LU}}$. This allows us to compute the position of the barycenter B, which serves as the center of the CR3BP coordinate system. 

We will start by finding a relation between the E point (CR3BP Earth) and the B point (CR3BP Barycenter) from the barycentre equality:
\begin{align}
    &m_{\text{M}} \Vec{BM} + m_{\text{E}} \Vec{BE} = \Vec{0},\\
    &\Vec{ME} = \Vec{MB} + \Vec{BE} = \left(1+\frac{m_{\text{M}}}{m_{\text{E}}}\right) \Vec{MB} = \frac{1}{1-\mu}\Vec{MB},
\end{align} 
where M is the Moon center, $m_{\text{X}}$ is the mass of body X and $\mu = \frac{m_{\text{M}}}{m_{\text{M}}+m_{\text{E}}}$ is the ratio of mass. Therefore we can compute the position of the B point:
\begin{align}
    \Vec{MB} = (1-\mu)\Vec{ME} = (1-\mu)R_{\text{LU}}\frac{\Vec{ME}_\text{eph}}{|\Vec{ME}_\text{eph}|},
\end{align} 
where the subscript ``eph" stands for ephemeris. Another fitting strategy is to consider a factor of $\frac{|\Vec{ME}|}{R_{\text{LU}}}$ to scale the spacecraft's state. According to our experimental results in this work, the first fitting strategy in Equations (1-3) yields better performance. Hence, only the average Earth-Moon distance is used without scalling. Lastly, we logically fit the rotational axes on the $u_{\text{R}}$, $u_\theta$ and $u_{\text{Z}}$ derived from the ephemeris Earth position with respect to the Moon:
\[
\begin{aligned}
    &\vec{u}_{\text{R}} = -\frac{\vec{ME}}{|\vec{ME}|}\\
    &\vec{u}_\theta = -\frac{\vec{V}}{|\vec{V}|} \hspace{\parindent}\text{with } \vec{V} = \frac{d\vec{ME}}{dt}\\
    &\vec{u}_{\text{Z}} = \vec{u}_{\text{R}} \times \vec{u}_\theta
\end{aligned}
\]

Now that the models are fitted, we can proceed to the conversion:
\begin{align}
    \Vec{MS}_{[\text{I}]} &= \vec{MB}_{[\text{I}]} + \vec{BS}_{[\text{I}]}\nonumber\\
    &= \vec{MB}_{[\text{I}]} + M_{\text{RI}} \cdot \vec{BS}_{[\text{R}]}\nonumber\\
    &= \vec{MB}_{[\text{I}]} + X_{[\text{R}]}\cdot\vec{u}_{\text{R}} + Y_{[\text{R}]}.\Vec{u}_\theta + Z_{[\text{R}]}.\vec{u}_{\text{Z}}\\
    \nonumber
    \Vec{V}_{\text{S/M[I]}} &= \Vec{V}_{\text{S/B[I]}} + \Vec{V}_{\text{B/M[I]}}\nonumber\\
    &= \Vec{V}_{\text{S/B[RI]}} + \Vec{\Omega}_{\text{R/I}} \times \Vec{BS}_{[\text{I}]} + \Vec{V}_{\text{B/M[I]}}\nonumber\\
    &=  M_{\text{RI}} \cdot \Vec{V}_{\text{S/B[R]}} + \Vec{\Omega}_{\text{R/I}} \times \Vec{BS}_{[\text{I}]} + \Vec{V}_{\text{B/M[I]}}
\end{align}
And $\Vec{\Omega}_{\text{R/I}}$ is defined as follows:
\begin{align}
    &\Vec{\Omega}_{\text{R/I}} = \Omega_{\text{R/I}} \cdot \vec{u}_{\text{Z}} \label{4},\\
    &\frac{d\vec{u}_{\text{R}}}{dt} = \Omega_{\text{R/I}} \cdot \Vec{u}_\theta \label{5},
\end{align}
In the equations above, M indicates the Moon, B the Barycenter, S the Spacecraft, [I] the inertial frame J2000, [R] the rotational frame, [RI] the frame with the same center as [R] but same axes as [I], $M_{\text{RI}}$ the transition matrix from [R] to [I], ($X,Y,Z$) the coordinates of $\vec{BS}$, $\Vec{V}_{\text{S/M[I]}}$ the velocity of S with respect to M in [I] and $\Vec{\Omega}_{\text{R/I}}$ the rotational velocity between the two frames (expressed in [I]), defined in Equations \ref{4} and \ref{5}. Equation \ref{4} defines the vector's direction and Equation \ref{5} defines the norm from the rotational unit vectors. 

We can then calculate the state in the inertial frame from the rotational one, with the previous conversion, and we can apply this process to our initial condition for the propagator. As the fitting cannot be perfect, starting from a DRO or NRHO trajectory, the ephemeris trajectory is not a real DRO or NRHO anymore, but we have a good initial guess close to the solution to initiate a convergence algorithm. See Section \ref{3BPOA} for further information. 

We consider the second case now where we want to convert the trajectory from the inertial frame to the rotational one. To convert the whole trajectory, we have to apply this conversion to every point of the trajectory.
\begin{align}
    \Vec{MS}_{[\text{R}]} &= M_{\text{IR}}\cdot\Vec{MS}_{[\text{I}]}\nonumber\\
    &= M_{\text{RI}}^{-1}\cdot\Vec{MS}_{[\text{I}]}\\
    \nonumber
    \Vec{V}_{\text{S/M[R]}} &= M_{\text{IR}}\cdot\Vec{V}_{\text{S/M[RI]}}\nonumber\\
    &= M_{\text{RI}}^{-1}\cdot\left(\Vec{V}_{\text{S/M[I]}} - \vec{\Omega}_{\text{R/I}}\Vec{MS}_{[\text{I}]} \right)
\end{align}
where $M_{\text{RI}}$ is defined by $u_{\text{R}}$, $u_\theta$ and $u_{\text{Z}}$ in column, similarly as in the first case. 

\section{High-fidelity ephemeris models} \label{Modelling}
The cis-lunar domain is influenced by multiple forces, with their impact varying depending on a spacecraft's position. Certain orbits of interest, such as those described by the TBP, exist due to the delicate balance between the gravitational pulls of the Earth and the Moon. To study these orbits, it is necessary to develop a high-precision propagator, and in this section, all relevant forces in the cis-lunar domain will be modeled. Implementation considerations will also be discussed to ensure the accuracy of HALO. The goal of this section is to provide a detailed understanding of the propagator's inner workings, facilitating its use in future studies. The propagator presented here builds on the work of Ennio Condoleo \citep{InitProp}, which has been further refined, detailed, and integrated into a mission design tool. This section will present the modelling for all field and surface forces and will then present some useful implementation consideration.

\subsection{Modelling of field forces}
\subsubsection{Lunar gravitational field} \label{LGF}
The assumption of a homogeneous LGF is incorrect. Therefore, the point mass model cannot be used. The Moon contains multiple ''mascons" (mass concentrations) unevenly distributed due to its geological history \citep{Mascons}. Consequently, the Moon's gravitational attraction varies with the relative position around the body. The gravitational filed potential $U$ is typically modeled in the body-fixed frame aligned with the Moon's principal axes (Moon PA frame), using an expansion in spherical harmonics (see \citep{SatOrbMod}, Page 56):
\begin{equation}
U = \frac{\mu}{r}\sum_{n=0}^{\infty} \sum_{m=0}^n\frac{R^n}{r^n}P_{nm}(\sin\phi)(C_{nm}\cos(m\lambda)+S_{nm}\sin(m\lambda)),
\end{equation}
where $\mu$ is the gravitational constant of the Moon and $R$ is its average radius. The coefficients $C_{nm}$ and $S_{nm}$ are the spherical harmonics coefficients, and can be calculated via the following expressions:
\[
\begin{aligned}
C_{n m} & =\frac{2-\delta_{0 m}}{M} \frac{(n-m)!}{(n+m)!} \int \frac{s^n}{R^n} P_{n m}\left(\sin \phi^{\prime}\right) \cos \left(m \lambda^{\prime}\right) \rho(s) d^3 s, \\
S_{n m} & =\frac{2-\delta_{0 m}}{M} \frac{(n-m)!}{(n+m)!} \int \frac{s^n}{R^n} P_{n m}\left(\sin \phi^{\prime}\right) \sin \left(m \lambda^{\prime}\right) \rho(s) d^3 s,
\end{aligned}
\]
with $P_{nm}$ being the Legendre polynomials, $\delta$ being the Kronecker's symbol (1 if both index are the same, else 0) and $\rho$ being the mass density at a given point. $(r,\phi,\lambda)$ and $(s,\phi',\lambda')$ correspond to the selenodetic coordinates, i.e., radius, latitude, longitude of the spacecraft and Moon's mass point, respectively.

Subsequently, the gravitational acceleration can be computed via the gradient of the potential. The coefficient $C_{nm}$ and $S_{nm}$ are processed during precise mission aiming at mapping the lunar geopotential like the GRAIL mission for the Moon \citep{Grail}. 
The implementation of this force is further discussed in Section \ref{Implem}.

\subsubsection{Lunar solid tides}
Solid tides, which affect solid materials, apply to any celestial body, including the Moon. The Moon's rotation period is almost exactly equal to its orbital period around the Earth, causing the same side of the Moon to always face the Earth. Consequently, the solid tides exerted by the Earth on the Moon are consistently directed in the same way within the Moon's body-fixed frame. They become a static problem that can be factored into the calculations of the LGF parameters $C_{nm}$ and $S_{nm}$. 
Therefore, it is crucial to use the most accurate data for these coefficients to achieve the highest precision in modeling both the LGF and the lunar solid tides.
\subsubsection{Point mass attraction} \label{PointMass}
When we consider orbits in the vicinity of the Moon, the bodies with the most influence are the Moon obviously, followed by the Earth, the Sun and then Jupiter and Venus. Even though Jupiter is further than Venus, its huge mass creates a bigger attraction on lunar orbits.

\begin{figure}[!ht]
    \centering
    \includegraphics[width=0.5\linewidth]{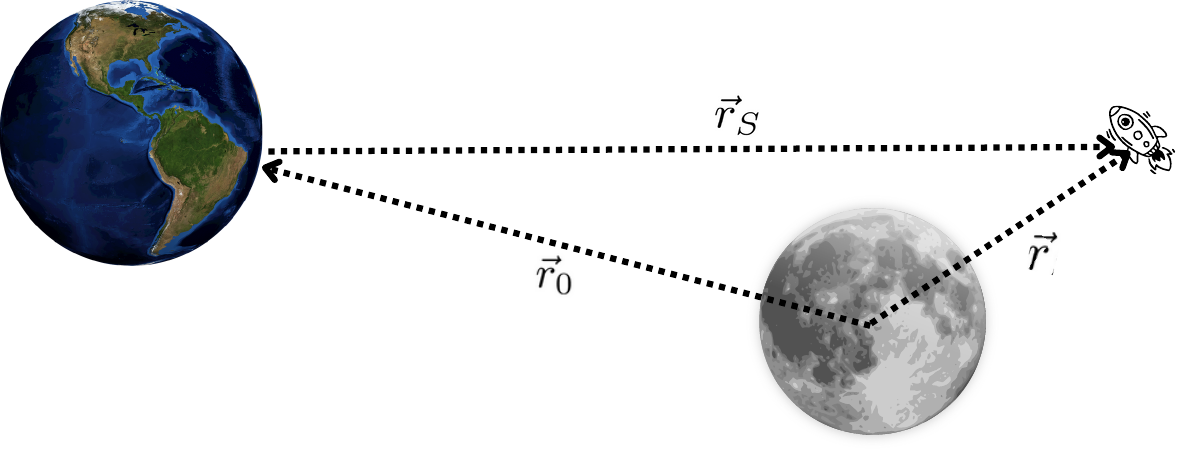}
    \caption{Third body perturbation - vectors definition}
    \label{fig:TBP}
\end{figure}

These bodies are considered as point masses when calculating their gravitational pulls. However, as opposed to the Moon, those additional bodies $\text{C}$ are third body perturbation and they have an attraction on both the spacecraft S and the Moon M. The resulting perturbation can be written:
\begin{align}
\Vec{a}_{\text{C/S [M]}} &= \Vec{a}_{\text{C/S [J2000]}} - \Vec{a}_{\text{C/M [J2000]}} \nonumber\\
&= -\mu \frac{\Vec{r}_{\text{S}}}{r_{\text{S}}^3} - \mu\frac{\Vec{r}_0}{r_0^3}, \label{PM perturbation}
\end{align}
where $\Vec{r}_{\text{S}}$ represent the vector from Earth to the spacecraft, and $\vec{r}_0$ represent the vector from the Moon to Earth. The notation “C/S” stands for “Body over Spacecraft”. The geometry of Earth-Moon-Spacecraft is illustrated in Figure \ref{fig:TBP}. Equation \ref{PM perturbation} indicates that the third body perturbation on the spacecraft corresponds to the attraction force of the body on the spacecraft subtracted by the attraction force on the Moon.

\subsubsection{Earth gravitational attraction}
To consider the Earth gravitational field (EGF) in a Moon centered frame, the following point mass model in Equation \ref{eq_point_mass_earth_moon} needs to be replaced by a harmonic model in Equation \ref{eq_harmonics_model_earth_moon}:
\begin{align}
\Vec{a}_{\text{E/S [M]}} = -\mu\frac{\Vec{r}_{\text{S}}}{r_{\text{S}}^3} -\mu\frac{\Vec{r_0}}{r_0^3}
\label{eq_point_mass_earth_moon}
\end{align}
\begin{align}
\Vec{a}_{\text{E/S [M]}} = A_{\text{EGF}}(\Vec{r}_{\text{S}}) - A_{\text{EGF}}(-\Vec{r_0})
\label{eq_harmonics_model_earth_moon}
\end{align}
where $r_{\text{S}}$ is the Earth-Spacecraft radial vector, $r_0$ is the Moon-Earth radial vector, and $A_{\text{EGF}}$ is the lunar gravitational model developed in Section \ref{LGF} but with Earth's coefficients. Hence the two terms on the right hand side of Equation \ref{eq_harmonics_model_earth_moon} can be computed as the LGF but with different $C_{nm}$ and $S_{nm}$ coefficients associated to the Earth.

\subsubsection{General relativistic correction}
The special relativity theory, that we usually use, considers a flat spacetime. However when the spacecraft is in the vicinity of a massive body, one has to take into account corrections related to the general relativity, as the spacetime is bent. Indeed, the acceleration correction due to the general relativity $\Vec{a}_{\text{rel}}$ can be expressed as in \citep{SatOrbMod} from Page 110:
\begin{align}
\Vec{a}_{\text{rel}}=\frac{G M}{r^2}\left(\left(4 \frac{G M}{c^2 r}-\frac{v^2}{c^2}\right) \Vec{e_r}+4 \frac{v^2}{c^2}\left(\Vec{e_r} \cdot \Vec{e_v}\right) \Vec{e_v}\right)
\end{align}
where $c$ is the speed of light, $\Vec{e_r}$ is the unit vector in the direction of position vector $\vec{r}$ and $\vec{e_v}$ is the unit vector in the direction of velocity vector $\vec{v}$.

\subsection{Modelling of surface forces}

\subsubsection{Solar radiation pressure} \label{SRP}
The solar radiations creates a pressure on the spacecraft, which can be modeled as(\citep{SatOrbMod} Page 77):
\begin{align}
P = \frac{L_{\text{s}}}{4\pi c R_{\text{S}}^2},
\end{align}
where $L_{\text{s}}$ is the Sun brightness power and $R_{\text{S}}$ is the distance to the Sun. Assume a spherical shaped spacecraft with only the specular reflections, the acceleration due to the solar radiation pressue is given by:
\begin{align}
\Vec{a}_{\text{SRP}} = \frac{PAC_{\text{R}}}{m}\Vec{e}_S     
\end{align}
where $\Vec{e}_S$ represents the unit vector from the Sun to the spacecraft; the cross section of the spacecraft $A$ is assumed circular and $C_{\text{R}}$ is the radiation pressure coefficient. $C_{\text{R}}$ lies between 0 and 2, with 0 accounting for no reflection, 1 for a black body and 2 for a total reflection.

\begin{figure}
    \centering
    \includegraphics[width=0.5\linewidth]{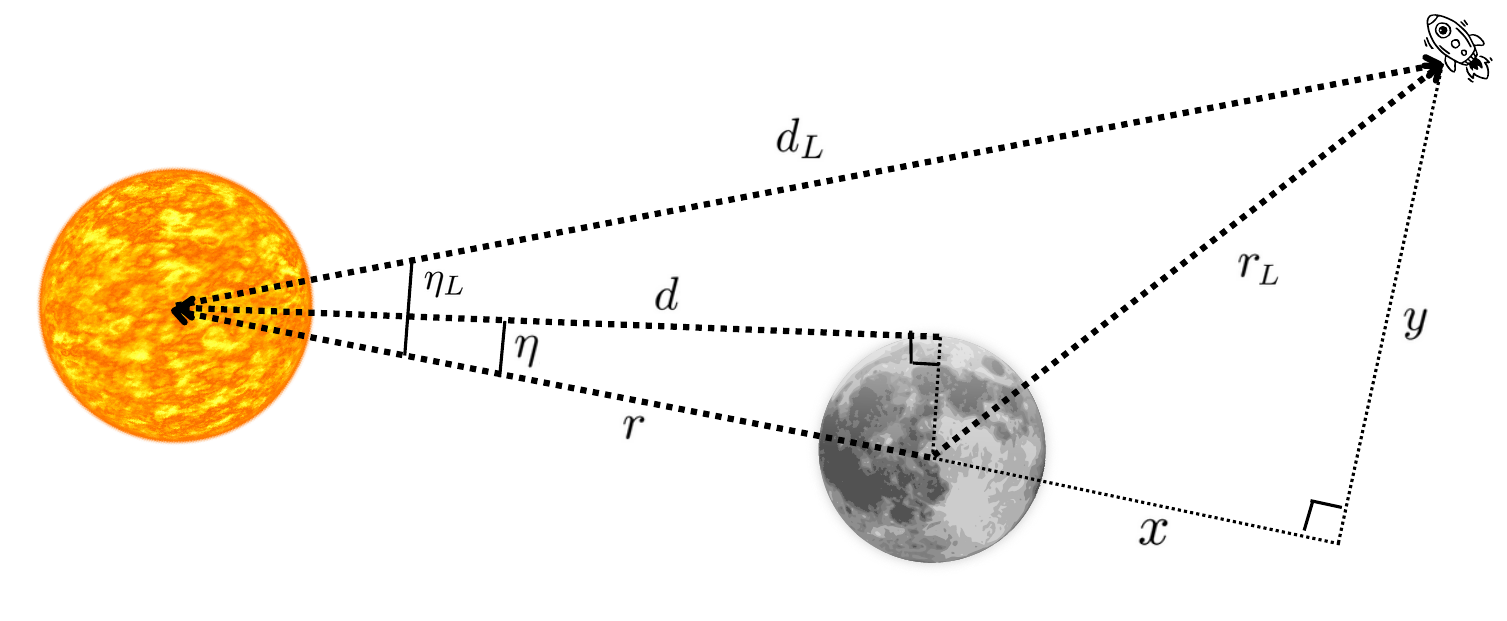}
    \caption{Eclipse condition}
    \label{fig:EclCond}
\end{figure}

An eclipse model should be considered when the Sun does not illuminate the spacecraft, which means the spacecraft is behind the Moon. All notations are defined in Figure \ref{fig:EclCond}. The condition of being hidden by the Moon is:
\begin{align}
    \left\{
    \begin{tabular}{c}
         $|\eta_{\text{L}}| < |\eta|$ \\
         $d_{\text{L}}>d$
    \end{tabular}
    \right.
    \Leftrightarrow 
    \left\{
    \begin{tabular}{c}
         $\cos(\eta_{\text{L}})>\cos(\eta)$ \\
         $d_{\text{L}}>d$
    \end{tabular}
    \right. \label{eclipse}
\end{align}
And with geometrical considerations (see Figure \ref{fig:EclCond}), we have:
\begin{align}
&\cos(\eta) = \frac{d}{r}\label{ceta} \\
&\cos(\eta_{\text{L}}) = \frac{r+x}{d_{\text{L}}} \nonumber\\
& x^2 + y^2 = r_L^2 \nonumber\\
& (x+r)^2 + y^2 = d_{\text{L}}^2 \nonumber
\end{align}
\text{Consequently, $x$ is solved as:}
\begin{align*}
    x = \frac{d_{\text{L}}^2 - r_L^2 -r^2}{2r}
\end{align*}
\text{Finally, $\cos(\eta_{\text{L}})$ is:}
\begin{align}
    \cos(\eta_{\text{L}}) = (r+\frac{d_{\text{L}}^2 - r_L^2 -r^2}{2r})/d_{\text{L}} = \frac{d_{\text{L}}^2 - r_L^2 + r^2}{2r d_{\text{L}}} \label{cetaL}
\end{align}
Thus Equations \ref{ceta} and \ref{cetaL} express the variables in Equation \ref{eclipse} and the eclipse condition is obtained from geometrical considerations.

\subsubsection{Earth albedo}
The bond Earth albedo corresponds to the electromagnetic energy reflected by the Earth from the Sun. We can apply the same eclipse condition for Earth this time, where the radiation pressure received by the Earth is:
\begin{align}
P_{\text{E}} = \frac{L_{\text{s}}}{4\pi r_{\text{S}}^2c}
\end{align}
where $r_{\text{S}}$ is the Earth-Sun distance, with the other notations remaining the same.
The solar flux reaches Earth surface over an area of $\pi R_{\text{T}}^2$. When average over a full day, the effective flux received by Earth is reduced to $P_{\text{E}}/4$ as measured at a distance $R_{\text{T}}$ from the Earth's center. Taking into account the inverse square law and considering Earth's albedo coefficient $C_{\text{alb}}$, the pressure exerted on the spacecraft is then given by:
\begin{align}
P = \frac{P_{\text{E}}}{4}C_{\text{alb}} \left(\frac{R_{\text{T}}}{d_{\text{L}}}\right)^2
\end{align}
wtih $d_{\text{L}}$ being the Earth-Spacecraft distance .

This model for albedo is really simple and lots of elements are not taken into account as the Earth area elements or angles to normal or cloud coverage, see \citep{SatOrbMod} for more information. But the resultant acceleration is already really low as we can see in the forces summary subsection.

\subsection{Considerations on implementation} \label{Implem}

\subsubsection{Implementation of the Lunar gravitational field}
The coefficients are usually normalized (we add a bar on top: $\bar{C}_{nm}$) in order to keep the magnitudes from changing significantly, so we have to convert the equations: 
\begin{equation}
\left\{\begin{array}{c}
\bar{C}_{n m} \\
\bar{S}_{n m}
\end{array}\right\}=\sqrt{\frac{(n+m)!}{\left(2-\delta_{0 m}\right)(2 n+1)(n-m)!}}\left\{\begin{array}{c}
C_{n m} \\
S_{n m}
\end{array}\right\}
\label{BarCoef}
\end{equation}

Thus we have also to consider the new resulting Legendre polynomial and the resulting form for the acceleration due to the LGF:

\begin{align}
&\bar{P}_{n m}=\sqrt{\frac{\left(2-\delta_{0 m}\right)(2 n+1)(n-m)!}{(n+m)!}} P_{n m} \label{Polynome}\\
&\ddot{\boldsymbol{r}}=\nabla \frac{\mu}{r} \sum_{n=0}^{\infty} \sum_{m=0}^n \frac{R^n}{r^n} \bar{P}_{n m}(\sin \phi)\left(\bar{C}_{n m} \cos (m \lambda)+\bar{S}_{n m} \sin (m \lambda)\right)
\end{align}

To compute this acceleration, we have to consider two things, computing the Legendre polynomials efficiently and deriving the gradient in Cartesian coordinate from the spherical ones.

First, in source \citep{SatOrbMod} on Page 66, we find recursion formulas to generate all the Legendre polynomials. However, we use normalised polynomials, so we need to perform a conversion\footnote{Additionally, in the function accelharmonic.m of the HALO tool \citep{HALO}, we use a different standard for the polynomial indices, where $n_C = n+1$ and the same for $m_C$ (C for ``Code"). The indices are shifted up by one, so the equations after conversion differ by one index from the equations in the code. The same adjustment is made for the coefficients $C_{nm}$ and $S_{nm}$ because MATLAB table indices start at 1, not 0.}. Therefore:
\begin{align}
\bar{P}_{n n} &= \sqrt{\frac{\left(2-\delta_{0 n}\right)(2 n+1)(n-n)!}{(2n)!}} P_{n n} \label{UBC1}\\
&=\sqrt{\frac{2\cdot(2 n+1)}{(2n)!}}\cdot(2n-1)\cos(\phi)P_{n-1,n-1} \label{Recursion}\\
&=\sqrt{\frac{2\cdot(2 n+1)}{(2n)!}}\cdot(2n-1)\cos(\phi)\cdot\bar{P}_{n-1,n-1}\cdot\sqrt{\frac{(2n-2)!}{2\cdot(2 n-1)}} \label{bar}\\
&=\sqrt{\frac{(2 n+1)(2n-1)}{(2n-1)2n}}\cdot\cos(\phi)\cdot\bar{P}_{n-1,n-1} \nonumber\\
&=\sqrt{\frac{(2 n+1)}{2n}}\cdot\cos(\phi)\cdot\bar{P}_{n-1,n-1} \label{Final}
\end{align}

We used in Equations \ref{UBC1} and \ref{bar} the two statements: $2-\delta_{0 n} = 2$ and $2-\delta_{0,n-1} = 2$ which are only valid for $n>1$. The first terms of $P_{nn}$ are then initialised from the value in \citep{SatOrbMod}: $P_{00} = 1$ and $P_{11} = \sqrt{3}\cos\phi$.

Equation \ref{UBC1} is derived from Equation \ref{Polynome}, and Equation \ref{Recursion} is based on the recursive Equation 3.23 from \citep{SatOrbMod}. Using Equation \ref{bar}, we revert to the ``bar" polynomial. Finally, with Equation \ref{Final}, we simplify the equation in a manner similar to Equations \ref{9} and \ref{10} for the other cases.
\begin{align}
\bar{P}_{n,n-1} &= \sqrt{\frac{2(2 n+1)1}{(2n-1)!}} \cdot (2n-1)\sin(\phi)\bar{P}_{n-1,n-1}\cdot 1/\sqrt{\frac{2(2 n-1)1}{(2n-2)!}} \nonumber\\
&= \sqrt{\frac{(2 n+1)}{(2n-1)(2n-1)}} \cdot (2n-1)\sin(\phi)\bar{P}_{n-1,n-1} \nonumber\\
&= (2n+1)\sin(\phi)\bar{P}_{n-1,n-1} \label{9}\\
\bar{P}_{n,m} &= \sqrt{\frac{2(2 n+1)(n-m)!}{(n+m)!}} \cdot \frac{1}{n-m}\left( (2n-1)\sin(\phi)\bar{P}_{n-1,m}\sqrt{\frac{(n+m-1)!}{2(2 n-1)(n-m-1)!}} ... \right. \nonumber\\
& \left. \hspace{\parindent}- (n+m-1)\bar{P}_{n-2,m}\cdot\sqrt{\frac{(n+m-2)!}{2(2 n-3)(n-m-2)!}} \right) \nonumber\\
&= \frac{1}{n-m}\left( (2n-1)\sin(\phi)\bar{P}_{n-1,m}\sqrt{\frac{2(2n+1)(n-m)!(n+m-1)!}{(n+m)!\cdot2(2 n-1)(n-m-1)!}} ... \right. \nonumber\\
& \left. \hspace{\parindent}- (n+m-1)\bar{P}_{n-2,m}\cdot\sqrt{\frac{2(2n+1)(n-m)!(n+m-2)!}{(n+m)!\cdot2(2 n-3)(n-m-2)!}} \right) \nonumber\\
&= \frac{1}{n-m}\left( (2n-1)\sin(\phi)\bar{P}_{n-1,m}\sqrt{\frac{(2n+1)(n-m)}{(n+m)(2 n-1)}} ... \right. \nonumber\\
& \left. \hspace{\parindent}- (n+m-1)\bar{P}_{n-2,m}\cdot\sqrt{\frac{(2n+1)(n-m-1)(n-m)}{(2 n-3)(n+m-1)(n+m)}} \right) \nonumber\\
&= \sin(\phi)\bar{P}_{n-1,m}\sqrt{\frac{(2n+1)(2 n-1)}{(n+m)(n-m)}} - \bar{P}_{n-2,m}\cdot\sqrt{\frac{(2n+1)(n-m-1)(n+m-1)}{(2 n-3)(n+m)(n-m)}} \nonumber\\
&= \sqrt{\frac{2n+1}{(n-m)(n+m)}}\left( \sin(\phi)\bar{P}_{n-1,m}\sqrt{2 n-1} - \bar{P}_{n-2,m}\cdot\sqrt{\frac{(n-m-1)(n+m-1)}{(2 n-3)}} \right) \label{10}
\end{align}

In addition, the derivative of the Legendre polynomials with respect to $\phi$ is computed in the same time from the obtained equations which is needed in the computation of the spherical gradient. 

With those 3 equations we can compute all Legendre polynomials. If we consider the coefficient in a triangle ($n,m\le n$), the first equation allows to compute the diagonal ($n,n$), then the second equation allows to compute the second diagonal (just under, ($n,n-1$)) from the first one, and the last formula allows to compute any other term from the 2 above itself.

Finally, the gradient is computed. However, the formulas for the gravity potential are in the spherical coordinates and the propagator needs Cartesian input and output. So first, the conversion from Cartesian to spherical coordinates for the spacecraft is done and then the gradient is computed with spherical coordinates as follows\footnote{Equations \ref{12}, \ref{13}, \ref{14} can be found in the the function accelharmonic.m of the HALO tool \citep{HALO}.}:
\begin{align}
&\nabla U = \frac{\partial U}{\partial r} \hat{\mathbf{r}}+\frac{1}{r} \frac{\partial U}{\partial \phi} \hat{\boldsymbol{\phi}}+\frac{1}{r \cos \phi} \frac{\partial U}{\partial \lambda} \hat{\boldsymbol{\lambda}}
\end{align}
With the following terms:
\begin{align}
\frac{\partial U}{\partial r} &= \frac{\partial}{\partial r} \left(\frac{G M}{r} \sum_{n=0}^{\infty} \sum_{m=0}^n \frac{R^n}{r^n} \bar{P}_{n m}(\sin \phi)\left(\bar{C}_{n m} \cos (m \lambda)+\bar{S}_{n m} \sin (m \lambda)\right)\right)  \nonumber\\
&= G M \sum_{n=0}^{\infty} \sum_{m=0}^n \frac{\partial}{\partial r} \left(\frac{R^n}{r^{n+1}}\right) \bar{P}_{n m}(\sin \phi)\left(\bar{C}_{n m} \cos (m \lambda)+\bar{S}_{n m} \sin (m \lambda)\right)  \nonumber\\
&= -(n+1)G M \sum_{n=0}^{\infty} \sum_{m=0}^n \frac{R^n}{r^{n+2}} \bar{P}_{n m}(\sin \phi)\left(\bar{C}_{n m} \cos (m \lambda)+\bar{S}_{n m} \sin (m \lambda)\right) \label{12}\\
\frac{\partial U}{\partial \phi} &= \frac{\partial}{\partial \phi} \left(\frac{G M}{r} \sum_{n=0}^{\infty} \sum_{m=0}^n \frac{R^n}{r^n} \bar{P}_{n m}(\sin \phi)\left(\bar{C}_{n m} \cos (m \lambda)+\bar{S}_{n m} \sin (m \lambda)\right)\right)  \nonumber\\
 &= \frac{G M}{r} \sum_{n=0}^{\infty} \sum_{m=0}^n \frac{R^n}{r^n} \frac{\partial}{\partial \phi} \left(\bar{P}_{n m}(\sin \phi)\right)\left(\bar{C}_{n m} \cos (m \lambda)+\bar{S}_{n m} \sin (m \lambda)\right) \label{13}\\
\frac{\partial U}{\partial \lambda} &= \frac{\partial}{\partial \phi} \left(\frac{G M}{r} \sum_{n=0}^{\infty} \sum_{m=0}^n \frac{R^n}{r^n} \bar{P}_{n m}(\sin \phi)\left(\bar{C}_{n m} \cos (m \lambda)+\bar{S}_{n m} \sin (m \lambda)\right)\right)  \nonumber\\
 &= \frac{G M}{r} \sum_{n=0}^{\infty} \sum_{m=0}^n \frac{R^n}{r^n} m\bar{P}_{n m}(\sin \phi)\left(-\bar{C}_{n m} \sin (m \lambda)+\bar{S}_{n m} \cos (m \lambda)\right) \label{14}
\end{align}
Here the vector form $\hat{\boldsymbol{r}}$, $\hat{\boldsymbol{\phi}}$, $\hat{\boldsymbol{\lambda}}$ corresponds to the spherical unit vectors. Then in the following, $\hat{\boldsymbol{x}}$, $\hat{\boldsymbol{y}}$, $\hat{\boldsymbol{z}}$ correspond to the Cartesian unit vectors. 

After that, we can convert back the computed gradient to obtain the Cartesian acceleration.
For that, we can use geometrical considerations on the sphere or use a formulary for the conversion from spherical unit vectors to cartesian unit vectors\footnote{Depending on the standard, formulas can be in radius $r$, latitude $\phi$ and longitude $\lambda$, or in radius $r$, polar angle $\theta = \phi - \frac{\pi}{2}$ and azimuthal angle $\varphi=\lambda$. On unit vectors, the only change is then $\hat{\boldsymbol{\theta}} = -\hat{\boldsymbol{\phi}}$.}:
\begin{align}
\hat{\mathbf{r}} & =\frac{x \hat{\mathbf{x}}+y \hat{\mathbf{y}}+z \hat{\mathbf{z}}}{\sqrt{x^2+y^2+z^2}} =\frac{x \hat{\mathbf{x}}+y \hat{\mathbf{y}}+z \hat{\mathbf{z}}}{r}  \nonumber\\
\hat{\boldsymbol{\phi}} & =\frac{\left(x^2+y^2\right) \hat{\mathbf{z}}-(x \hat{\mathbf{x}}+y \hat{\mathbf{y}}) z}{\sqrt{x^2+y^2+z^2} \sqrt{x^2+y^2}}=\frac{\left(x^2+y^2\right) \hat{\mathbf{z}}-(x \hat{\mathbf{x}}+y \hat{\mathbf{y}}) z}{r \sqrt{x^2+y^2}}  \nonumber\\
\hat{\boldsymbol{\lambda}} & =\frac{-y \hat{\mathbf{x}}+x \hat{\mathbf{y}}}{\sqrt{x^2+y^2}} \nonumber
\end{align}

And with geometrical consideration on the sphere, by considering the circular cut of the three dimensional sphere, parallel to the x,y plane and going through the point of interest, we have:
\begin{align}
r\cos\phi &= \sqrt{x^2+y^2} \hspace*{\fill} \notag
\end{align}
We can now calculate the gradient and convert it to Cartesian coordinates:
\begin{align}
\hspace{\parindent}\nabla U &= \frac{\partial U}{\partial r} \hat{\mathbf{r}}+\frac{1}{r} \frac{\partial U}{\partial \phi} \hat{\boldsymbol{\phi}}+\frac{1}{\sqrt{x^2+y^2}} \frac{\partial U}{\partial \lambda} \hat{\boldsymbol{\lambda}} \hspace{\parindent} \nonumber\\
&= \frac{\partial U}{\partial r} \frac{x \hat{\mathbf{x}}+y \hat{\mathbf{y}}+z \hat{\mathbf{z}}}{r}+\frac{1}{r} \frac{\partial U}{\partial \phi} \frac{\left(x^2+y^2\right) \hat{\mathbf{z}}-(x \hat{\mathbf{x}}+y \hat{\mathbf{y}}) z}{r \sqrt{x^2+y^2}}+\frac{1}{\sqrt{x^2+y^2}} \frac{\partial U}{\partial \lambda} \frac{-y \hat{\mathbf{x}}+x \hat{\mathbf{y}}}{\sqrt{x^2+y^2}} \nonumber\\
& = \hspace{10px} \left[ \frac{x}{r} \frac{\partial U}{\partial r} - \frac{xz}{r^2 \sqrt{x^2+y^2}}\frac{\partial U}{\partial \phi} - \frac{y}{x^2+y^2} \frac{\partial U}{\partial \lambda} \right] \hat{\mathbf{x}} \label{15}\\
& \hspace{11px} + \left[ \frac{y}{r} \frac{\partial U}{\partial r} - \frac{yz}{r^2 \sqrt{x^2+y^2}}\frac{\partial U}{\partial \phi} + \frac{x}{x^2+y^2} \frac{\partial U}{\partial \lambda} \right] \hat{\mathbf{y}} \nonumber\\
& \hspace{11px} + \left[ \frac{z}{r} \frac{\partial U}{\partial r} + \frac{\sqrt{x^2+y^2}}{r^2}\frac{\partial U}{\partial \phi} \right] \hat{\mathbf{z}} \nonumber
\end{align}

According to Equation \ref{15}, the acceleration due to the LGF is expressed with respect to the three Cartesian coordinates in the Moon PA frame.

Therefore this harmonic form allows us to consider a more precise gravitational filed than the point mass. In order to be the most accurate possible on the LGF, which is the dominant force, and on the lunar solid tides, we need the most accurate data for the $C_{nm}$ and $S_{nm}$ coefficients. The most accurate mission in this domain was the GRAIL mission \citep{Grail} in 2012 with two spacecrafts GRAIL A/B working in pair to precisely compute the $C_{nm}$ and $S_{nm}$ coefficients. Those coefficients were then implemented into the propagator with a gravitational field model going up to $350\times350$ harmonics.

\subsubsection{Lunar gravitational field assessment}

Now we want to study the accuracy introduced by those new harmonics by using the example of an LLO orbit. According to Figure \ref{fig:PartAcc}, we have that $a_{\text{LGF}} = 1.418763\cdot10^{-3}$ km/s² and the Earth albedo effect, yielding the lowest perturbation, is on the order $10^{-16}$. The norm of the LGF for different harmonic numbers is further investigated across various harmonic orders/degrees. The difference compared to the previous harmonics is computed.

\begin{table}[!ht]
    \centering
    \begin{tabular}{|c||c|c|c|c|c|c|c|c|}
         \hline
         Harmonic orders & (0,0) & (2,2) & (10,10) & (30,30) & (70,70)\\
         \hline \hline
         Norm of LGF (km/s²) & 1.417777e-03 & 1.418120e-03 & 1.418604e-03 & 1.418674e-03 & 1.418763e-03\\
         \hline
         Improvement & \O & 3.4231e-07 & 4.8433e-07 & 6.9740e-08 & 8.9285e-08\\
         \hline \hline
         Harmonic orders & (100,100) & (150,150) & (200,200) & (250,250) & (350,350)\\
         \hline \hline
         Norm of LGF (km/s²) & 1.418759e-03 & 1.418759e-03 & 1.418759e-03 & 1.418759e-03 & 1.418759e-03\\
         \hline
         Improvement & 3.7071e-09 & 4.6582e-10 & 7.9744e-12 & 9.3311e-14 & 1.1055e-14\\
         \hline
    \end{tabular}
    \caption{Improvement of the LGF with respect to the orders of harmonics}
    \label{tab:LGF}
\end{table}
In Table \ref{tab:LGF}, we observe that the error associated with $70\times70$ harmonics is on the order of $10^{-9}$ (sum of the remaining errors), while for $150\times150$ harmonics, the error decreases to $10^{-11}$. Considering the norms of all other forces and the computational times—27.0 seconds for the $70\times70$ configuration and 66.5 seconds for the $150\times150$ configuration in LLO\footnote{This computational time is provided for reference and was obtained using a middle-range Intel Core i5 11$^{th}$ generation computer.}—we will adopt the $150\times150$ harmonics for all propagations in this study. Utilizing a higher number of harmonics would result in excessive computational demands. Although this conclusion is derived from a single orbit, we assume its general applicability.

\subsubsection{Earth gravitational field assessment}
Similarly to the LGF, we have selected the EGM2008 model \citep{pavlis2012development} for Earth coefficients, which is widely used for the EGF. We have conducted a comparable study on the harmonics accuracy as we did for the LGF. Considering the LLO orbit, we have $a_{\text{EGF}} = 1.8465\cdot 10^{-8}$ km/s², while the albedo effect, representing the lowest perturbation, is on the order of $10^{-16}$.

We will analyze the differences across various harmonic numbers between the norm of the EGF and the norm of the point mass Earth attraction (EPM). This analysis will illustrate the enhancement of each harmonic relative to the reference Earth attraction. Furthermore, we will quantify the improvement in accuracy compared to preceding harmonics.

\begin{table}[!ht]
    \centering
    \begin{tabular}{|c||c|c|c|c|c|}
         \hline
         Harmonic numbers & (0,0) & (2,2) & (3,3) & (10,10) & (100,100)  \\
         \hline \hline
         $|\Vec{a}_{\text{EGF}} - \Vec{a}_{\text{EPM}}| (km/s^2)$ & 1.0641e-21 & 1.6227e-14 & 1.6230e-14 & 1.6230e-14 & 1.6230e-14 \\
         \hline
         Improvement & \O & 1.6227e-14 & 3e-18 & $<$ 1e-19 & $<<$ 1e-19\\
         \hline
    \end{tabular}
    \caption{Improvement of the EGF with respect to the number of harmonics}
    \label{tab:EGF}
\end{table}

As shown in Table \ref{tab:EGF}, the $0\times0$ harmonic exhibits an order of $10^{-21}$, indicating that the EGF aligns with the point mass EGF (EPM), as anticipated. The improvement of the $3\times3$ harmonic is lower than the lowest force taken into account (order of $10^{-16}$). Moreover, the computational time for the $0\times0$ and $3\times3$ harmonics is roughly the same in LLO, ranging from 65 to 69 seconds. Therefore, there is no need to use orders higher than $3\times3$. Similar to the findings for the LGF, this result, although derived from a single orbit, is expected to have broader applicability.

\subsubsection{Integration settings}
In addition to force modelling and perturbation effects, the choice of integrator and its parameters is crucial to the propagator's accuracy, particularly for certain orbits. This section presents a comparative analysis of different integrators.

The Lunar Reconnaissance Orbiter (LRO) spacecraft, orbiting in LLO, was propagated for four days, with position outputs every two hours. This allowed us to assess orbital evolution errors over multiple revolutions. Additionally, the choice of integrator has a significant impact on the accuracy of LLO orbit propagation (see Section \ref{Sum Up Pert}), making it a focal point for this study.

For each integrator, the position root mean square error (RMSE) over the four-day propagation was compared with varying tolerances. Two types of tolerances were examined: relative and absolute. Relative tolerance measures error relative to the magnitude of each solution component, while absolute tolerance defines a threshold below which the value of a solution component is considered insignificant. In our analysis, absolute tolerance did not affect accuracy and was fixed at $10^{-14}$, leaving relative tolerance as the primary variable. The default relative tolerance was $10^{-6}$, as per \citep{InitProp}.

Given the need for high-accuracy integration in this study, we tested various integrators from MATLAB's ODE suite \citep{ODE}. We selected four integrators for comparison: MATLAB ODE45, ODE78, ODE89, and ODE113. The first three are Runge-Kutta (RK) methods, where ``ODE” is followed by two numbers, $n$ and $m$ ($m > n$), representing the order $m$ of the integrator, with errors proportional to $h^{m+1}$. The integration step size is adjusted based on the computational error, derived from the difference between orders $n$ and $m$. ODE45 uses the Dormand-Prince RK45 formula \citep{ODE45}, while ODE78 and ODE89 are based on Verner's RK formulas, with ODE89 being more robust and ODE78 more efficient \citep{ODE789}. ODE113 is a variable-step, variable-order method using the Adams-Bashforth-Moulton algorithm, with orders ranging from 1 to 13 \citep{ODE113}.

The following table presents the RMSE for the four integrators used to propagate the LRO on February 1, 2020. In Table \ref{tab:Integ}, the integrator combination from the initial study is highlighted in blue. As evident from the table, this choice results in a significant RMSE for this specific orbit. The four values highlighted in red represent the three lowest RMSEs across all integrators, as well as the lowest value achieved by ODE45.

\begin{table}[!ht]
    \centering
    \begin{tabular}{|c||c|c|c|c|c|}
        \hline
        Relative tolerance, $10^{-n}$ & 6 & 7 & 8 & 9 & 11\\
        \hline \hline
        ODE45 RMSE (m) & \textcolor{blue}{26758} & 645.0 & 729.8 &  \textcolor{red}{421.1} & 422.2 \\
        Computing time (s) & 19.4 & 36.4 & 63.3 & 97.7 & 236.3\\
        \hline
        ODE78 RMSE (m) & 5255 & 20901 & 1092 & 381.9 & 422.0\\
        Computing time (s) & 29.4 & 44.8 & 66.8 & 97.2 & 190.2\\
        \hline
        ODE89 RMSE (m) & \textcolor{red}{175.6} & \textcolor{red}{363.9} & 419.3 & 422.8 & 422.3\\
        Computing time (s) & 64.1 & 91.2 & 126.8 & 169.6 & 300.3\\
        \hline
        ODE113 RMSE (m) & 3150 & \textcolor{red}{291.9} & 418.9 & 425.3 & 422.4\\
        Computing time (s) & 40.8 & 64.0 & 92.9 & 129.5 & 206.4\\
        \hline
    \end{tabular}
    \caption{Integrators computing time and accuracies on February 1, 2020}
    \label{tab:Integ}
\end{table}

A key observation is that all integrators, when configured with sufficiently low tolerances, converge to a similar accuracy level, approximately 422 meters in this case. Specifically, ODE45 achieves this accuracy with an RT of $10^{-9}$, ODE78 with $10^{-11}$, ODE89 with $10^{-8}$, and ODE113 with $10^{-8}$. Among these, ODE45 and ODE113 are the fastest, reaching this accuracy in approximately 95 seconds. However, certain parameter combinations (shown in red) yield similar or even better accuracies in less computing time. To assess whether these low RMSEs are due to favorable integration errors or represent genuinely superior parameter choices, we will test these four red-highlighted combinations on a different epoch: February 6, 2020. The results are presented in Table \ref{tab:CSitu}.

\begin{table}[!ht]
    \centering
    \begin{tabular}{|c||c|c|}
        \hline
         & RMSE (m) & Computing time (s)\\
        \hline \hline
        ODE89, RT $10^{-6}$ & 682.5 & 66.3 \\
        \hline
        ODE89, RT $10^{-7}$ & 622.5 & 90.6 \\
        \hline
        ODE113, RT $10^{-7}$ & 269.6 & 62.1 \\
        \hline
        ODE45, RT $10^{-9}$ & 605.4 & 96.1 \\
        \hline
    \end{tabular}
    \caption{Results for the four selected combinations on February 6, 2020}
    \label{tab:CSitu}
\end{table}

In Table \ref{tab:CSitu}, the ODE45 result yields a stable final accuracy of 605.4 meters. Notably, the two ODE89 combinations approach this value but exceed it, indicating that their previously observed lower RMSEs in Table \ref{tab:Integ} were due to positive integration errors. However, ODE113 remains both the fastest and the most accurate integrator in this scenario. To further validate the choice of ODE113 over ODE45, we conducted an additional test on March 10, 2020.

\begin{table}[!ht]
    \centering
    \begin{tabular}{|c||c|c|}
        \hline
         & RMSE (m) & Computing time (s)\\
        \hline \hline
        ODE113, RT $10^{-7}$ & 147.6 & 67.4 \\
        \hline
        ODE45, RT $10^{-9}$ & 509.7 & 102.8 \\
        \hline
    \end{tabular}
    \caption{Results for the two selected comparisons on March 10, 2020}
    \label{tab:LSitu}
\end{table}

In this third case (Table \ref{tab:LSitu}), ODE113 continues to outperform, providing superior accuracy compared to the stable value achieved by other integrators with longer processing times. Nevertheless, while the ODE113 configuration shows excellent performance, the combination with RT=$10^{-7}$ exhibits less stability than the ODE113/RT=$10^{-8}$ configuration. This difference in stability is detailed in Table \ref{tab:Stab}.

\begin{table}[!ht]
    \centering
    \begin{tabular}{|c||c|c|}
        \hline
         & Reference RMSE (m) & w/o albedo RMSE (m)\\
        \hline \hline
        ODE113, RT $10^{-7}$, 01/02/2020 & 292 & 53 \\
        \hline
        ODE113, RT $10^{-7}$, 06/02/2020 & 270 & 399 \\
        \hline
        ODE113, RT $10^{-8}$, 01/02/2020 & 419 & 364 \\
        \hline
        ODE113, RT $10^{-8}$, 06/02/2020 & 566 & 595 \\
        \hline
    \end{tabular}
    \caption{Stability of the new combination}
    \label{tab:Stab}
\end{table}
The Earth albedo introduces a minimal perturbation in Low Lunar Orbit (LLO), with an order of magnitude of approximately $10^{-16}$, and its removal is not expected to significantly affect the results. However, with ODE113 at a relative tolerance (RT) of $10^{-7}$, while the RMSE is favorable, there is a notable variation in results when Earth albedo is removed, unlike the more stable behavior observed with ODE113 at RT=$10^{-8}$. This sensitivity can be attributed to the complexity of LLO and ELFO orbits, as discussed in Section \ref{Sum Up Pert}, where integration methods have a more pronounced impact.

Given these findings, ODE113 is selected as the preferred integrator due to its superior accuracy and computational efficiency. For complex orbits, such as LLO and ELFO, a relative tolerance of $10^{-8}$ will be used to ensure stability, while for less complex orbits, a tolerance of $10^{-7}$ will suffice.

\subsubsection{Time and coordinate systems handling}\label{sec:TimeCoord}
The mission design tool HALO uses NASA's SPICE system to manage reference frames, time conversions, and access ephemeris data for planets and spacecraft. SPICE, developed by the Navigation and Ancillary Information Facility (NAIF) \citep{SPICE}, supports multiple programming languages, including MATLAB and Python, through the Mice package and the spiceypy library, respectively. The same SPICE functions are used across both languages, with minor differences in function prefixes.

SPICE is primarily used for three key purposes:
\begin{enumerate}
    \item Time conversion: SPICE converts UTC to ET by accounting for leap seconds using time kernels (e.g., naif0012.tls). The function str2et is used in MATLAB, and datetime2et is used in Python.
    \item Ephemeris data access: To obtain planetary or spacecraft position data, SPICE loads binary ephemeris files (e.g., de430.bsp for planetary data). The function spkezr is used to retrieve the position of a target with respect to a reference frame.
    \item Frame transformations: SPICE computes transformations between reference frames using efficient matrix functions such as pxfrm2, sxform, and pxform.
\end{enumerate}

\subsection{Summary of all forces}
Here are then all the forces and perturbations taken into account by the HALO tool:
\begin{enumerate}
    \item Lunar gravitational field - $350\times350$ harmonics model from the GRAIL mission
    \item Earth gravitational field - $100\times100$ harmonics model EGM2008
    \item Sun and Jupiter point masses
    \item Solar radiation pressure with a spherical shape model
    \item General relativistic correction
    \item Earth albedo 
\end{enumerate}
In summary, this section details the development of the modeling framework utilized for our high-fidelity propagator. Some perturbations, such as lunar albedo and the gravitational influence of other planets in the solar system, were not further developed within this model due to their negligible impact relative to the primary perturbations considered. To enhance the accuracy of the propagator, the recommendations provided in the preceding relevant section or the general guidelines in Section \ref{Sum Up Pert} should be followed.

\section{Assessments using real-life cis-lunar orbits} \label{Assess}

\subsection{Process of assessments}
It is essential to assess whether modifications made to the propagator result in measurable improvements. To this end, we compare its propagation against real existing trajectories. This involves taking the real initial position of a spacecraft in an orbit of interest and propagating it, while comparing the propagated data to the real data every 30 seconds. The following four orbits are selected to assess the HALO tool: LLO, ELFO, NRHO, and DRO.

\subsection{Low Lunar Orbit}
\subsubsection{Reference orbit}
The LLO are commonly defined as the orbits under 100 km of the Moon's surface. These orbits are of significant interest for lunar exploration, as they enable detailed surface studies, such as mapping the gravitational field and analyzing lunar geology. Additionally, LLOs provide rapid access for potential landing missions.

One of the key challenges of LLOs is their high instability due to the Moon's complex gravitational field, primarily influenced by mass concentrations, or “mascons.” Without careful planning and regular adjustments, a spacecraft in LLO would eventually collide with the lunar surface within a few weeks. However, certain ``frozen orbits” exist, where gravitational perturbations are balanced, minimizing the need for frequent corrective maneuvers.

The LRO, launched in 2009, remains in a frozen polar orbit around the Moon, requiring only annual maneuvers to maintain its trajectory \citep{LRO}. This spacecraft serves as a reference for our LLO study, where we consider its orbit on February 1, 2020, over a four-day period. With an orbital period of approximately two hours, the LRO's periselene during this time is 60 km, and its aposelene is 130 km, averaging 95 km, which classifies it as a typical LLO. The data used was downloaded from the LRO Radio Science website\citep{LROData} with an orbit determination accuracy of ~10m.  With an integration time step of 30 seconds, the propagation consumes 64 seconds.

\begin{figure}[!ht]
    \centering
    \includegraphics[width=0.5\linewidth]{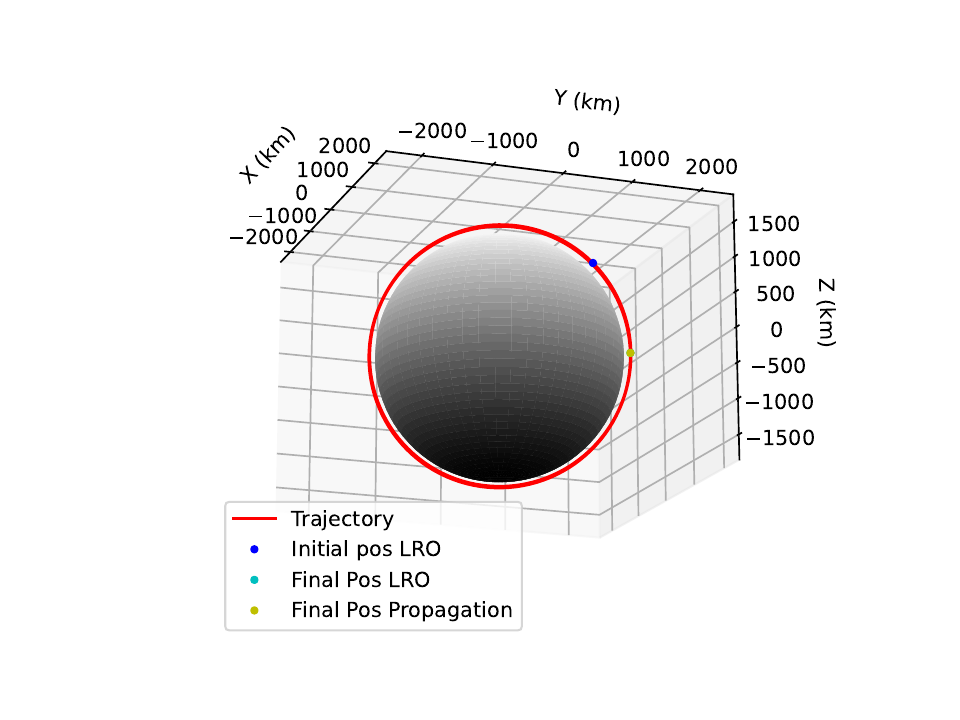}
    \caption{LRO Trajectory via orbit propagation}
    \label{fig:LRO_Prop}
\end{figure}

\begin{figure}[!ht]
    \centering
    \includegraphics[width=0.5\linewidth]{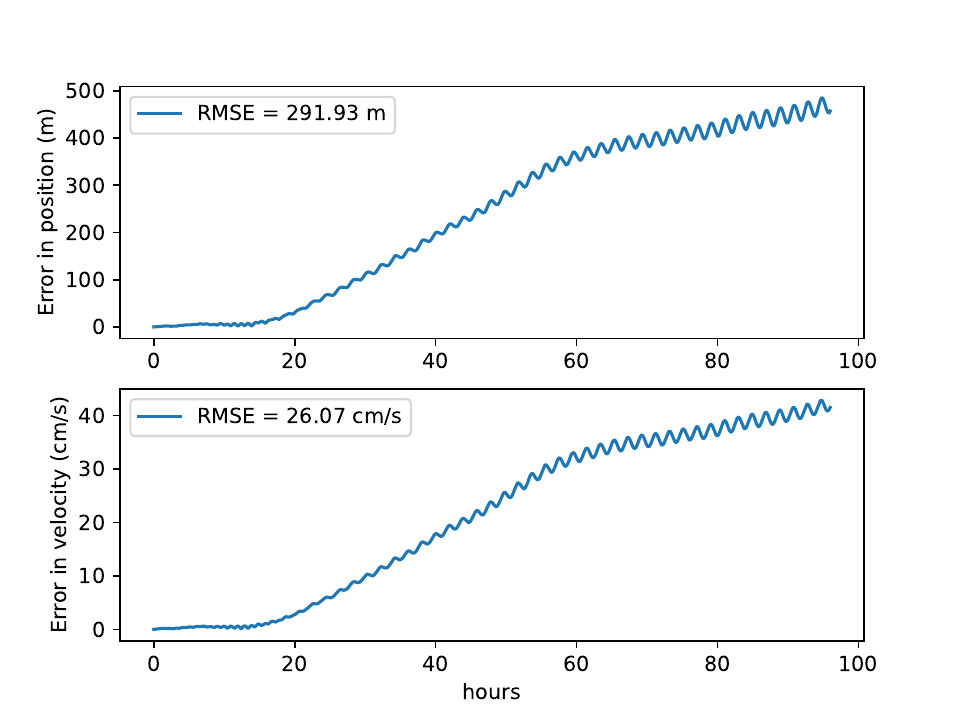}
    \caption{Propgation errors compared to the LRO reference orbit}
    \label{fig:LRO_Error}
\end{figure}

In Figure \ref{fig:LRO_Prop}, we cannot see the cyan dot because it is on the same location as the green one, and we can see on Figure \ref{fig:LRO_Error} that after 4 days the RMSE in position is 292 meters. This will be our reference for following modelling assessments for the propagator.

\subsubsection{Magnitude of each perturbation}

\begin{figure}[!ht]
    \centering
    \includegraphics[width=0.5\linewidth]{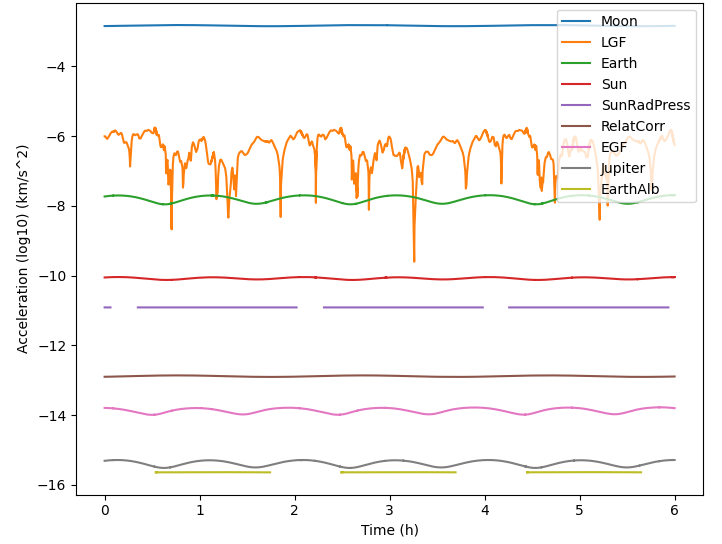}
    \caption{Magnitudes of perturbations for LRO in LLO}
    \label{fig:PartAcc}
\end{figure}

In Figure \ref{fig:PartAcc}, we can see all the perturbations applied to the spacecraft and their amplitudes. The span is 6 hours, corresponding to approximately 3 orbits in LLO. It is expected to see a predominance of the lunar gravitational attraction (labeled ``Moon" in the figure), which remains almost constant due to the relatively small amplitude of variation (only 70 km) compared to the Moon's radius.

The second predominant perturbation overall is the improvement of the LGF from a point mass model (labeled ``LGF" in the figure). This reflects the effects of the lunar ``mascons", which induce variations in gravitational pull, resulting in positional improvements that vary in magnitude. Notably, we observe a periodicity in the gravitational field improvements at approximately 2 hours and 4 hours, indicating similar trends during these time intervals.

Then the second most prominent body in this orbit is the Earth. We observe more fluctuations here than for the Moon even if the Earth is further away, it is due to the definition of the third body perturbation, see Section \ref{PointMass}. The Earth attraction is defined by a difference between the attraction on the spacecraft and the attraction on the Moon, so that when the spacecraft moves closer to the Moon, the difference tends to go to 0. We remark that the EGF improvement is way more stable than the LGF one because we only use 3 harmonics instead of 150 for the Moon.

Then we have the Sun attraction, the solar radiation pressure with some eclipse periods on the trajectory. The gaps in some functions are eclipse phenomena because the SRP and Earth albedo become null on specific occasion, see Section \ref{SRP}. Then really low, we have still the relativistic corrections, the EGF improvement, Jupiter's attraction and the Earth albedo with also eclipse periods.

This figure helps us to consider which perturbation are predominant on the orbit. The graph can help to put focus on the improvement of certain perturbations or to know which to remove without losing too much accuracy and then winning computational power.

\subsubsection{Sensitivity analysis}\label{RefLRO}

In this section, we systematically remove individual perturbations or improvements from the model to evaluate their relative impact on the propagator's performance, while keeping all other factors consistent with the reference scenario. As a reminder, the reference case involves a 4-day propagation of an LLO, with the following settings:  $150\times150$ LGF harmonics, $3\times3$ EGF harmonics, the Sun attraction, an SRP coefficient of 1.3, the GR correction, the Jupiter gravitational attraction, an Earth albedo coefficient of 0.3, the integrator ODE113 and a relative tolerance of $10^{-8}$.

\begin{table}[!ht]
    \centering
    \begin{tabular}{c||c|c|c|c|c|c}
        & Reference & $10\times10$ LGF & w/o Earth & w/o Sun & w/o SRP & w/o GRC \\
        RMSE (m) & 418.9 & 4548 & 814.7 & 387.6 & 377 & 389.2 \\
        \hline\hline
        & $0\times0$ EGF & w/o Jupiter & w/o Albedo & ODE45 (RT=$10^{-7}$) & RT of $10^{-6}$\\
        RMSE (m) & 408.5 & 441.2 & 363.7 & 602 & 2999
    \end{tabular}
    \caption{Relative role of each perturbation or improvement}
    \label{tab:relroleLLO}
\end{table}

Table \ref{tab:relroleLLO} shows that when removing perturbations with small impacts, the results remain close to the reference. For instance, removing the GRC, using a point mass model for Earth (i.e., $0\times0$ EGF), or excluding Jupiter's gravitational influence and Earth's albedo results in only slight deviations from the reference. These deviations are often slightly lower, which may be attributed to integration errors, as the orbit's complex dynamics can introduce such numerical inaccuracies. The two most critical parameters, are related to the integrator's settings, underscoring their importance for accurate LLO propagation. While Earth's albedo appears to be the smallest force in Figure \ref{tab:relroleLLO}, its 13\% deviation from the reference suggests that the observed difference may be a result of integration errors rather than the physical influence of the albedo itself.

Overall, the largest factor contributing to the accuracy is the inclusion of the LGF harmonics, followed by the integrator's relative tolerance and the Earth's attraction. While some values fall below the reference, which theoretically should not occur (as removing a perturbation should not improve accuracy), these discrepancies are likely due to integration errors rather than incorrect physical modeling. Given the relatively small differences, these results should be cross-checked with other orbital configurations for further validation.

\subsection{Elliptical Lunar Frozen Orbit}

\subsubsection{Reference orbit}
The ELFO is characterized by a highly eccentric orbit around the Moon, with a significant difference between perilune and apolune. The term ``frozen" refers to the orbit's stability over time, requiring minimal maintenance burns—a challenging task due to the Moon's complex gravitational field, similar to LLO. ELFOs are recognized by their periodicity in orbital element changes, ensuring long-term stability. 

JAXA has selected ELFO as the optimal orbit for the LNSS, a system currently under development for lunar exploration \citep{ELFO}. However, no spacecraft has yet flown in such orbits. To assess our propagator on a similar orbit, we focus on the Elliptical Polar Orbit (EPO) flown by the Clementine spacecraft, launched in January 1994. The data used was downloaded from the NAIF website\citep{SPICEData} but the orbit determination accuracy could not be found though being the best achieved for this orbit. Clementine entered an elliptical orbit around the Moon to conduct imaging and altimetry mapping between -60° and 60° latitude \citep{Clementine}. During its mission, two maneuvers were performed, with the final one on April 11, concluding the mapping phase by April 22. For our analysis, we selected the epoch of April 15, 1994, at 3 PM, which avoids maneuvers, and a 15-hour time span covering three periods of 5 hours each. This epoch and duration were chosen to avoid discontinuities in the orbit determination tracks, which occur approximately every 20 hours. The perilune for this orbit is 415 km, and the apolune is 2939 km. The propagation achieved a RMSE accuracy of 24.7 m in a computational time of 2.27 seconds, with an initial integration step size of 60 seconds.

\begin{figure}
    \centering
    \includegraphics[width=0.5\linewidth]{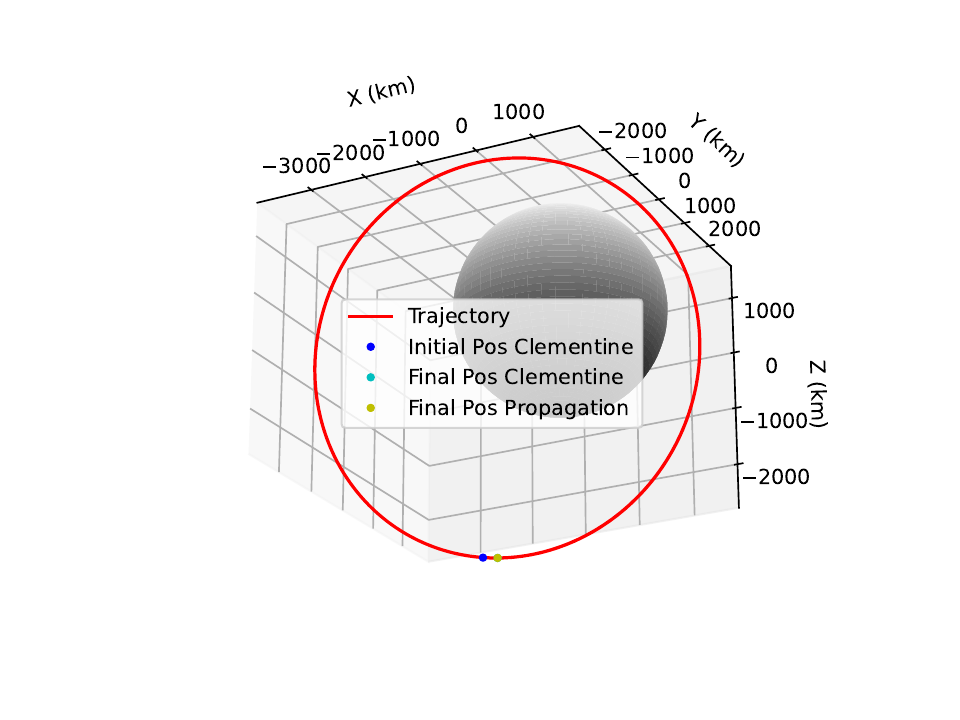}
    \caption{The Clementine spacecraft during its mapping mission}
    \label{fig:ELFO}
\end{figure}
\begin{figure}
    \centering
    \includegraphics[width=0.5\linewidth]{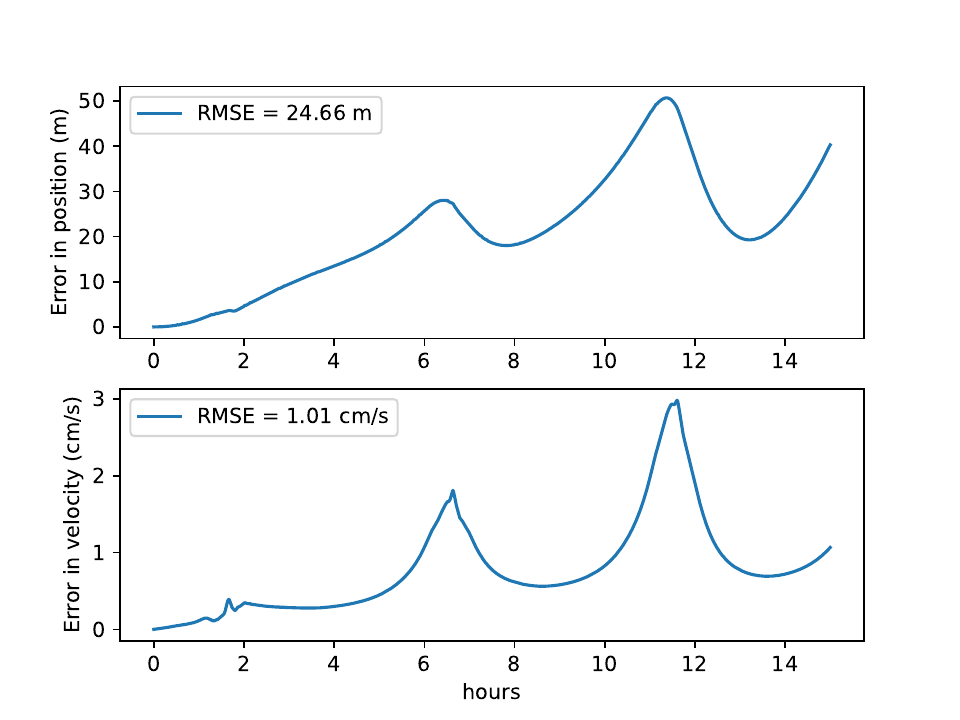}
    \caption{Propagation errors with Clementine}
    \label{fig:ELFOE}
\end{figure}

In the Figure \ref{fig:ELFO}, we can see the polar orbit of the Clementine spacecraft during the second phase of its mapping mission. The propagator performs with high accuracy over 15 hours, yielding a positioning error of around 20 m at apolune. Figure \ref{fig:ELFOE} shows a velocity error peak at perilune, where the spacecraft's velocity is the highest, leading to a corresponding increase in position error. This behavior is expected due to the geometry of the orbit.

\subsubsection{Magnitude of each perturbation}

\begin{figure}[!ht]
    \centering
    \includegraphics[width=0.5\linewidth]{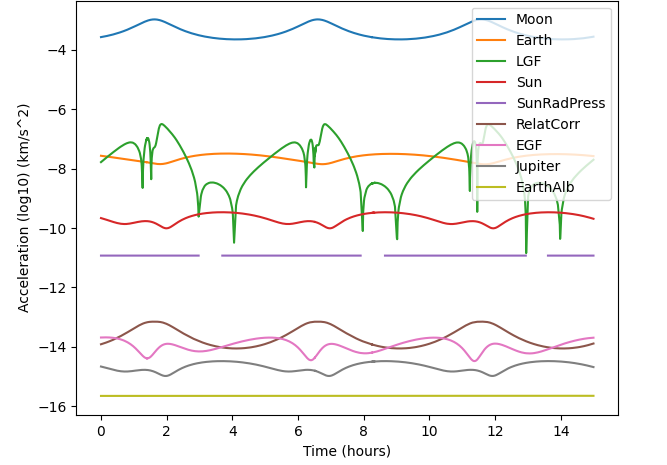}
    \caption{Magnitude of perturbations during a ELFO}
    \label{fig:PartAccELFO}
\end{figure}

In Figure \ref{fig:PartAccELFO}, we study the accelerations over 3 periods and we observe a periodic motion as expected. Despite significant altitude variations throughout the orbit, the spacecraft remains entirely within the Moon's SoI, which extends to around 65,000 km. As anticipated, a dominant force is the Moon attraction force, characterized by oscillations. The LGF improvements, modeled with 150 harmonics, also exhibit substantial variation due to the Moon's complex gravitational environment. We observe the Earth, Sun and SRP and the other perturbations are really small similarly to the LLO orbit. A notable feature is the inverse relationship between the third-body perturbations (from the Earth, Sun, and Jupiter) and the Moon's gravitational influence: third-body effects reach a minimum when the Moon's influence is at its peak. This behavior arises from the differential force calculation used for third-body perturbations, which approaches zero as the spacecraft nears the Moon. Additionally, eclipse periods are evident in the SRP data, occurring when the spacecraft is obscured by the Moon and thus shielded from direct sunlight. 
\subsubsection{Sensitivity analysis}

In this section, we will systematically remove each perturbation or improvement to assess their individual contributions to the propagator's accuracy, similar to the LLO. All other perturbations or improvements are kept consistent with the reference situation.

As a remainder, the reference model simulates an EPO over a 15-hour period, with the same parameters as the LLO discussed in Section \ref{RefLRO}, except for the relative tolerance, which is set to $10^{-8}$.

\begin{table}[!ht]
    \centering
    \begin{tabular}{c||c|c|c|c|c|c}
        & Reference & $10\times10$ LGF & w/o Earth & w/o Sun & w/o SRP & w/o GRC \\
        RMSE (m) & 24.7 & 61.0 & 2361 & 57.3 & 23.1 & 24.7 \\
        \hline\hline
        & $0\times0$ EGF & w/o Jupiter & w/o Albedo & ODE45 (RT=$10^{-7}$) & RT of $10^{-6}$\\
        RMSE (m) & 24.7 & 24.7 & 24.7 & 32.7 & 349
    \end{tabular}
    \caption{Relative role of each perturbation or improvement}
    \label{tab:relroleELFO}
\end{table}

Shown in Table \ref{tab:relroleELFO}, removing most perturbations generally decreases the model's accuracy, resulting in an increase in RMSE. An exception is observed when the SRP is removed, where the RMSE slightly decreases. This could be attributed to an integration error, similar to what was seen with the LLO, or it might indicate an issue with the SRP modeling itself, which will be discussed further in Section \ref{Sum Up Pert}.

From the results, it is clear that the Earth's gravitational influence and the relative tolerance have the most significant impact on this orbit, followed by the improvements to the LGF and the Sun's perturbation. This is consistent with the characteristics of the EPO, which is slightly farther from the Moon compared to an LLO, and thus more influenced by the Earth and the Sun. The sensitivity to the relative tolerance is likely due to the complexity of integrating this orbit, as previously discussed and further elaborated in Section \ref{Sum Up Pert}.

\subsection{Near Rectilinear Halo Orbit}
\subsubsection{Reference orbit} 
The NRHO has emerged as a prominent orbital configuration in recent years, serving as a focal point for future lunar missions and as a candidate orbit for space stations around the Moon. As a member of the halo orbit family, the NRHO is defined within the context of the TBP, distinguishing it from conventional elliptical orbits governed by the two-body problem. While spacecraft orbits around the Moon are often modeled using the two-body assumption, incorporating Earth's gravitational influence provides a more accurate analytic model. However, halo orbits are typically defined within the CR3BP, a simplified analytic framework that allows their efficient computation.

From the perspective of an Earth-based observer, a spacecraft in a halo orbit around the Moon would appear to exhibit a periodic motion resembling a “halo” around the Moon. This periodicity is defined in a Moon-centered rotating frame for the Earth-Moon system. A significant feature of halo orbits is their ability to maintain a relatively constant geometric relationship with both the Earth and the Moon, enabling continuous direct line-of-sight communication with Earth. This makes them particularly valuable for missions requiring sustained communication links.

The NRHO is a special type of halo orbit that brings the spacecraft closer to the Moon's surface, providing easier access to the lunar surface. However, NRHOs are not inherently stable and require periodic station-keeping maneuvers to maintain the orbit. Despite this, NRHOs have been chosen by NASA for its lunar space station Gateway \citep{Gateway}, which will use the NRHO for its planned orbital operations. NASA's CAPSTONE mission \citep{Capstone} was designed to explore and validate the characteristics of the selected NRHO, making CAPSTONE's trajectory a reference case for NRHO dynamics. The data used was downloaded from the Horizon website\citep{Horizon} but the orbit determination accuracy could not be found though being the best achieved for this orbit.

The NRHO flown by CAPSTONE exhibits a 9:2 synodic resonance, meaning the spacecraft completes nine orbits around the Moon while the Moon completes two orbits around the Earth. For this study, the chosen epoch is 25 November 2022, with a simulation span of 6.5 days, which approximates the average NRHO orbital period of 6.68 days. The orbit's perilune and apolune are 1,638 km and 69,083 km, respectively. The computational time for the orbit propagation, performed on a medium Intel i5 processor, is 1.2 seconds, yielding an RMSE of 1,443 meters. The initial integration step size before applying step size control was set to 60 seconds.

\begin{figure}
    \centering
    \includegraphics[width=0.7\linewidth]{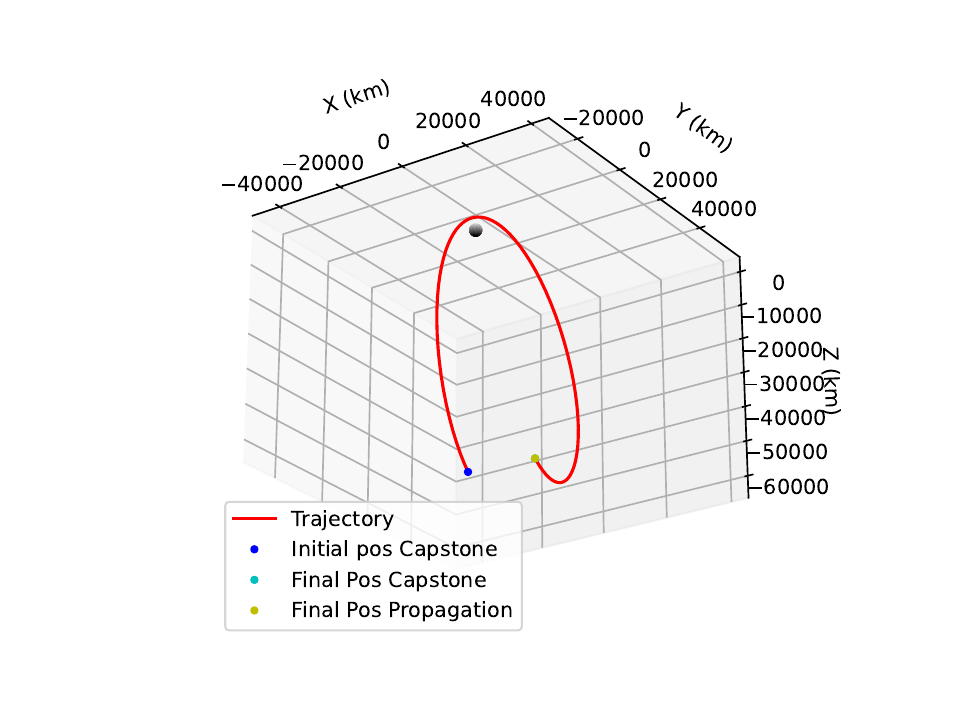}
    \caption{CAPSTONE trajectory in Moon centered initial frame}
    \label{fig:NRHOG}
\end{figure}
\begin{figure}
    \centering
    \includegraphics[width=0.7\linewidth]{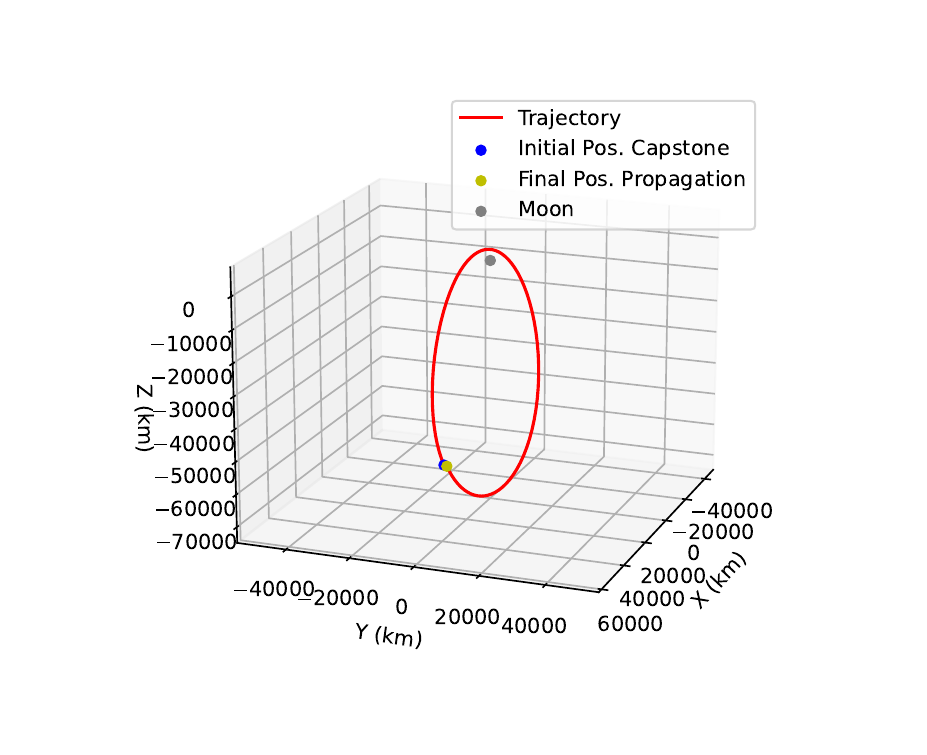}
    \caption{CAPSTONE trajectory in the Moon centered rotational frame}
    \label{fig:NRHOGR}
\end{figure}

\begin{figure}[!ht]
    \centering
    \includegraphics[width=0.5\linewidth]{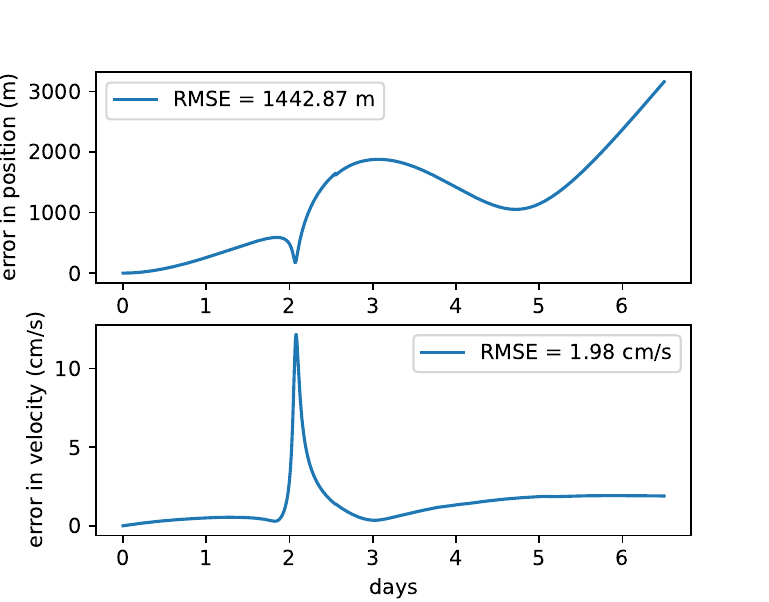}
    \caption{Propagation errors with CAPSTONE}
    \label{fig:NRHOE}
\end{figure}

In Figure \ref{fig:NRHOG}, the reference frame is the Moon-centered inertial frame (J2000), meaning the Earth revolves around the Moon, causing the spacecraft's orbit to appear non-closed. In this frame, the NRHO shows the relative motion of the spacecraft as both the Earth and Moon move. Conversely, in Figure \ref{fig:NRHOGR}, the orbit is shown in a Moon-centered rotating frame, where the Earth remains fixed at a position approximately -400,000 km away. In Figure \ref{fig:NRHOE}, we observe that the velocity error increases significantly near the Moon's surface. This increase in error is correlated with the higher absolute velocity of the spacecraft at perilune, which is expected given the orbital geometry. 

\subsubsection{Magnitude of each perturbation}

\begin{figure}[!ht]
    \centering
    \includegraphics[width=0.5\linewidth]{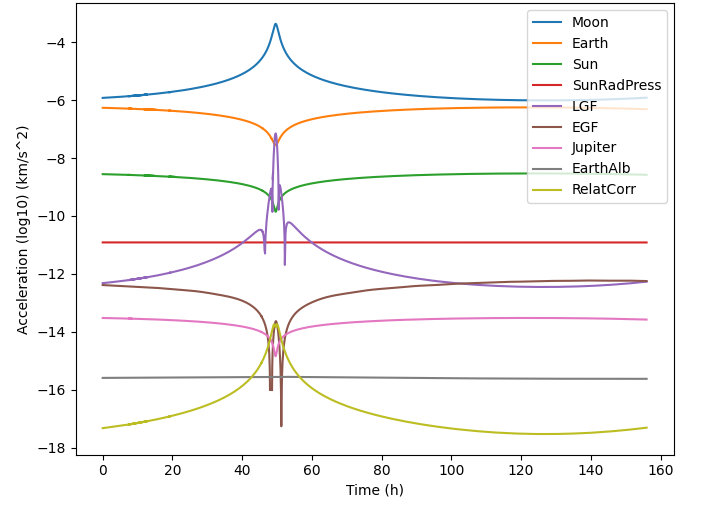}
    \caption{Magnitude of perturbations during a NRHO}
    \label{fig:PartAccNRHO}
\end{figure}

In Figure \ref{fig:PartAccNRHO}, we analyze the accelerations over one orbital period. As the spacecraft approaches perilune, we observe a sharp increase in lunar gravitational attraction, which peaks due to the spacecraft's proximity to the Moon. At apolune, approximately 70,000 km from the Moon's surface, the spacecraft is near the edge of the Moon's SoI, where the gravitational attractions of the Earth and Moon are almost equal. Close to the Moon, the perturbation due to Earth's gravity diminishes because the perturbation is defined as the difference between the accelerations acting on the spacecraft and the Moon. As the spacecraft nears the Moon, this difference becomes negligible, resulting in reduced Earth-attraction perturbation. The same effect is observed for the gravitational perturbations from the Sun and Jupiter.

The SRP remains relatively constant throughout the orbit, as it primarily depends on the distance to the Sun, which does not vary significantly. Similarly, the albedo effect from Earth remains steady, since the NRHO is nearly orthogonal to the Earth-Moon line, leading to minimal variation in the Earth-spacecraft distance.

For the gravitational filed improvements (LGF and EGF), it is observed that their impact follows the overall trend of the gravitational acceleration. The stronger the force, the more pronounced the improvement, with some irregularities attributed to the higher-order harmonics. However, the influence of the LGF is much lower in comparison to LLO, as expected for this orbital configuration.

Finally, the relativistic correction only becomes significant in regions where spacetime curvature is pronounced, such as near the Moon's surface. In this case, the last three perturbations—SRP, albedo, and relativistic effects—are largely negligible due to their minimal contributions to the spacecraft's overall motion.

\subsubsection{Sensitivity analysis}

In this section, we systematically remove each perturbation or improvement in the propagator to evaluate the relative contribution of each. All other perturbations and improvements remain consistent with the reference scenario.

As a reminder, the reference case involves an NRHO with a 6.5-day period, using the same model parameters as the LLO described in Section \ref{RefLRO}, except for the relative tolerance, which has been adjusted to $10^{-7}$.

\begin{table}[!ht]
    \centering
    \begin{tabular}{c||c|c|c|c|c|c}
        & Reference & $10\times10$ LGF & w/o Earth & w/o Sun & w/o SRP & w/o GRC \\
        RMSE (km) & 1.443 & 1.443 & 10 307 & 93.0 & 1.776 & 1.443 \\
        \hline\hline
        & $0\times0$ EGF & w/o Jupiter & w/o Albedo & ODE45 (RT=$10^{-7}$) & RT of $10^{-6}$\\
        RMSE (km) & 1.435 & 1.484 & 1.443 & 1.404 & 2.519
    \end{tabular}
    \caption{Relative role of each perturbation or improvement}
    \label{tab:relroleNRHO}
\end{table}

In Table \ref{tab:relroleNRHO}, the integrator's choice shows a slightly smaller impact on accuracy compared to the reference scenario, indicating that using ODE113 instead of ODE45 for the NRHO may not have been optimal. However, the overall effect is minor, with an error of less than 4\%, which suggests that this discrepancy may simply be a small integrator-related anomaly specific to this orbit.

The analysis shows that the most significant perturbations influencing the orbit are Earth's gravitational attraction, followed by the Sun and the SRP. In contrast, other perturbations have relatively minimal effects. As the spacecraft is further from the Moon compared to an LLO, the accuracy of the integrator's parameters becomes less critical, and the dominant factors become the gravitational influences of Earth and the Sun. This is consistent with expectations, as the spacecraft is near the edge of the Moon's SoI, where the Earth's gravitational impact becomes more pronounced.

\subsection{Distant Retrograde Orbit}

\subsubsection{Reference orbit}
The DRO is another key orbital configuration of the decade which attracts extensive research. Like the NRHO, the DRO is defined within the framework of the TBP.

When observed from Earth, a spacecraft in a DRO around the Moon exhibits a periodic motion, with the apparent geometry depending on the orbital period and resonant ratio. In a Moon-centered inertial frame, the DRO can trace shapes such as squares or triangles, while in a rotational frame, the orbit takes on a different, yet still closed, form. A defining characteristic of DROs is their retrograde nature, meaning that the spacecraft's rotation around the Moon is opposite to the direction of most celestial bodies in space.

DROs are known for their high stability, allowing spacecraft to remain in orbit for extended periods without the need for frequent maneuvers. NASA selected a DRO for the Artemis I mission, where the Orion spacecraft entered a 14-day DRO orbit for six days \citep{ArtemisI}. This spacecraft serves as the reference for our DRO assessment, with an epoch of November 29, 2022, at 4 PM, and a 24-hour span. The data used was downloaded from the Horizon website\citep{Horizon} but the orbit determination accuracy could not be found though being the best achieved for this orbit. The Artemis I mission's DRO has a 14-day period, with a perilune of 70,100 km and an apolune of 94,800 km. The computational time for the orbit propagation is 0.38 seconds, with an RMSE accuracy of 417.8 meters and an initial integration step size of 60 seconds.

\begin{figure}[!ht]
    \centering
    \includegraphics[width=0.7\linewidth]{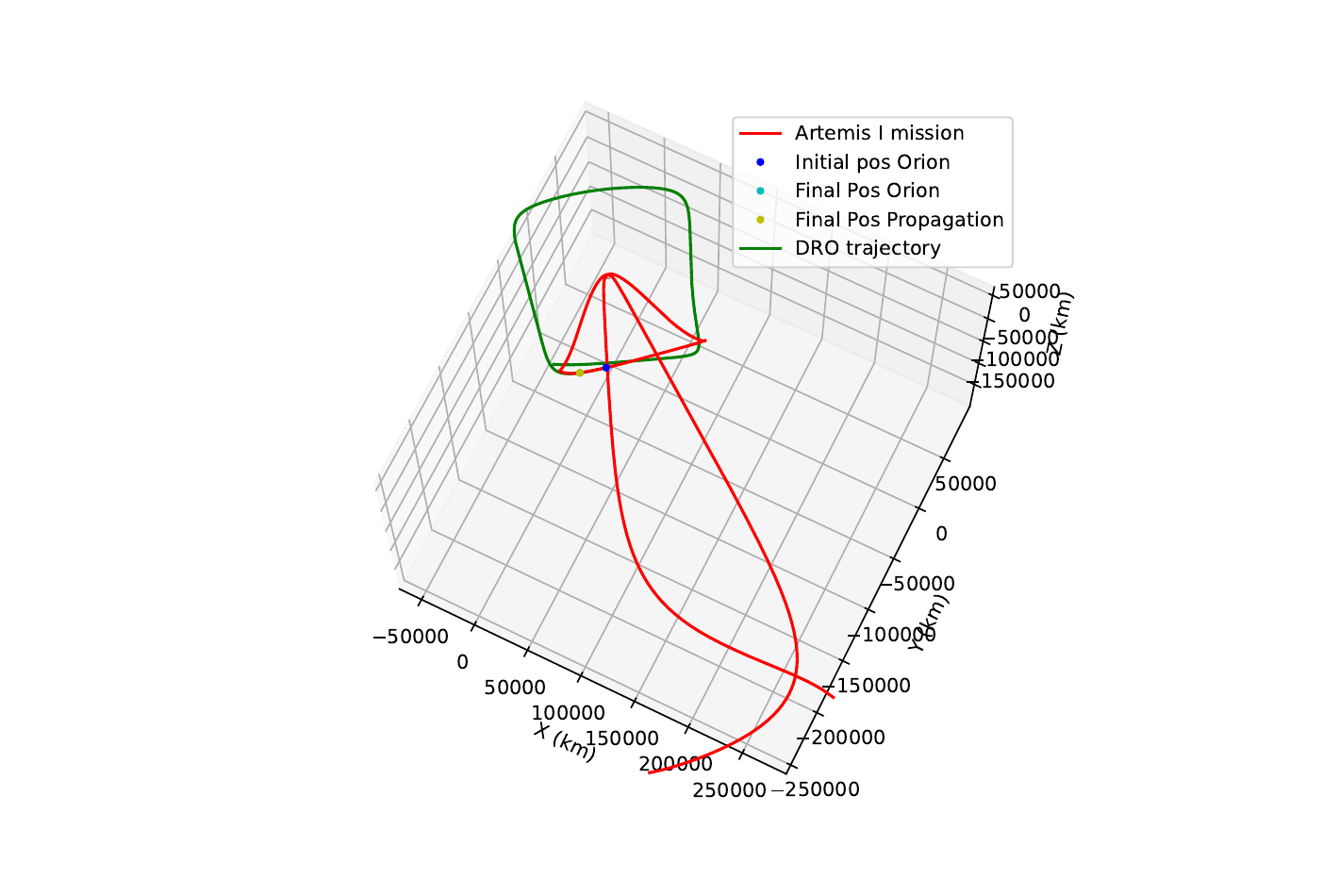}
    \caption{Artemis I mission in the inertial frame}
    \label{fig:Art1}
\end{figure}
\begin{figure}[!ht]
    \centering
    \includegraphics[width=0.5\linewidth]{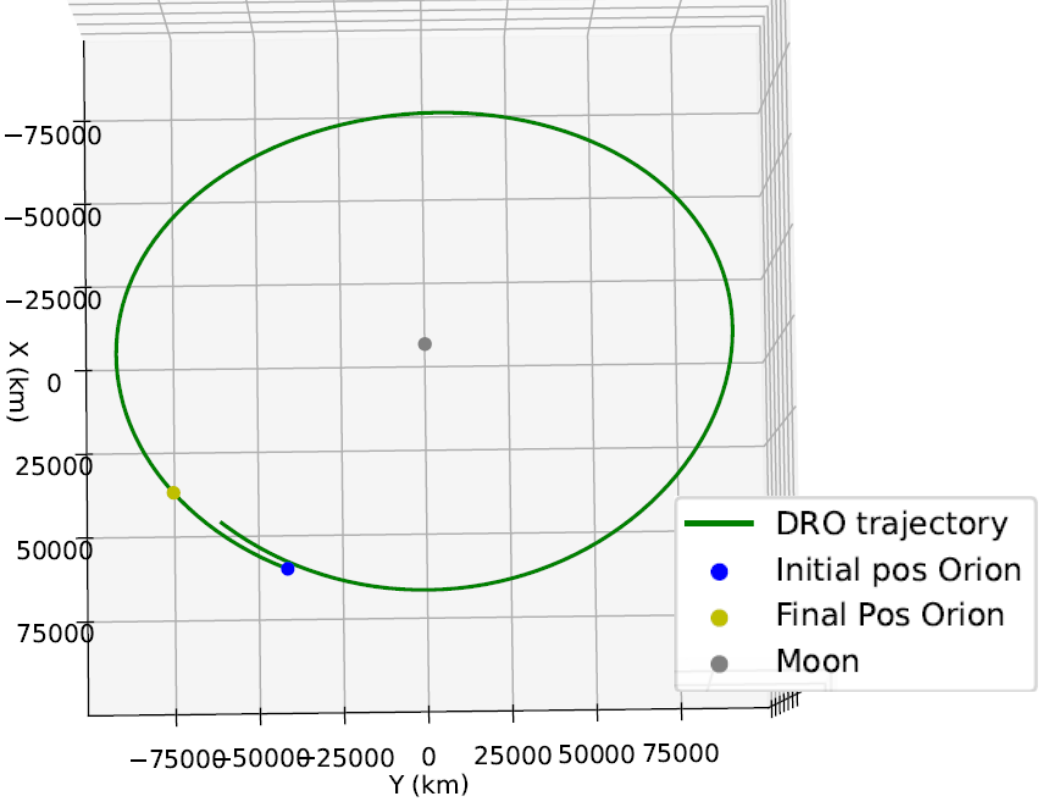}
    \caption{Orion's DRO orbit in the rotational frame (14 days)}
    \label{fig:DROR}
\end{figure}

\begin{figure}[!ht]
    \centering
    \includegraphics[width=0.5\linewidth]{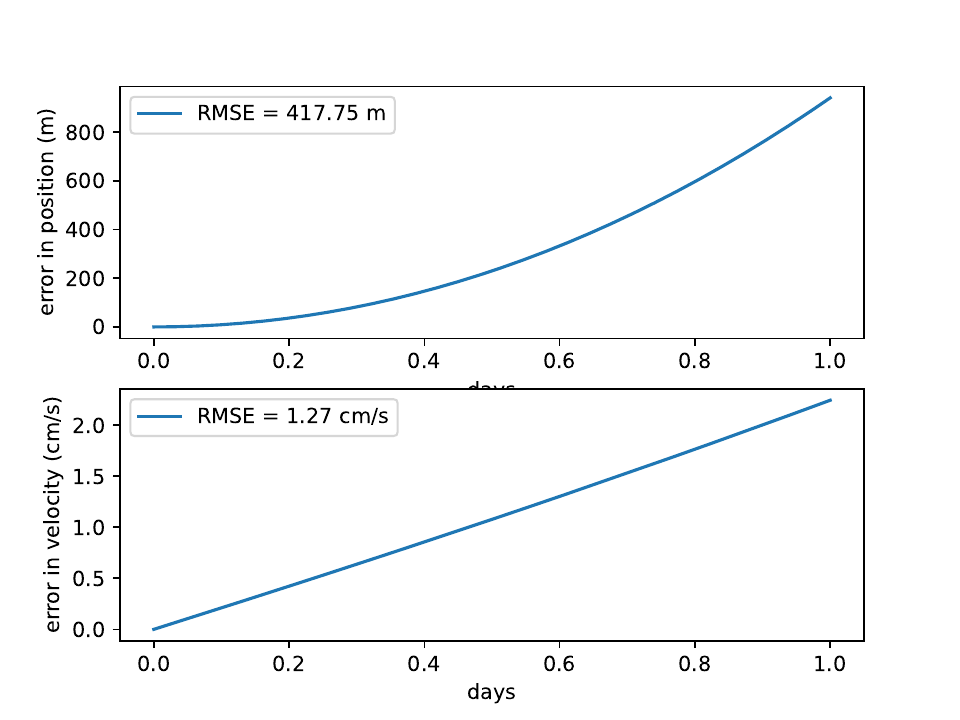}
    \caption{Propagation errors of the Orion}
    \label{fig:Art1E}
\end{figure}

In Figure \ref{fig:Art1}, we present a portion of the Artemis I mission trajectory in a Moon-centered inertial frame. This visualization focuses on a limited segment of the full trajectory, showing a small part of the DRO, covering only one day of the total six-day stay in DRO. The green line represents the DRO orbit that the Orion spacecraft followed before executing its exit maneuver. After 28 days (equivalent to two full orbital periods), the orbit forms an almost square shape. Figure \ref{fig:DROR} shows the Orion spacecraft's DRO in a Moon-centered rotational frame, where the orbit appears nearly closed. The plot clearly captures a period of just under 14 days, reflecting the periodic nature of the DRO. In Figure \ref{fig:Art1E}, we observe that the position error increases significantly at certain points, which may indicate limitations in the modeling for this specific orbit or part of the trajectory. This underscores the importance of evaluating the entire orbit over a complete revolution, which was not possible in this instance.

\subsubsection{Magnitude of each perturbation}

To assess the magnitude of each perturbation for the DRO, we cannot use the Orion spacecraft because the span is only 1 day over the 14-day period. Therefore, we took the associated orbit (square-shaped) and made it converge from the CR3BP model to the high-fidelity model that we use. This was done with the algorithm presented in Section \ref{3BPOA}. Then we studied the amplitude of all perturbations with this similar real DRO trajectory for half a square, which is 1 period, or 14 days (a full square is actually 2 periods).

\begin{figure}[!ht]
    \centering
    \includegraphics[width=0.5\linewidth]{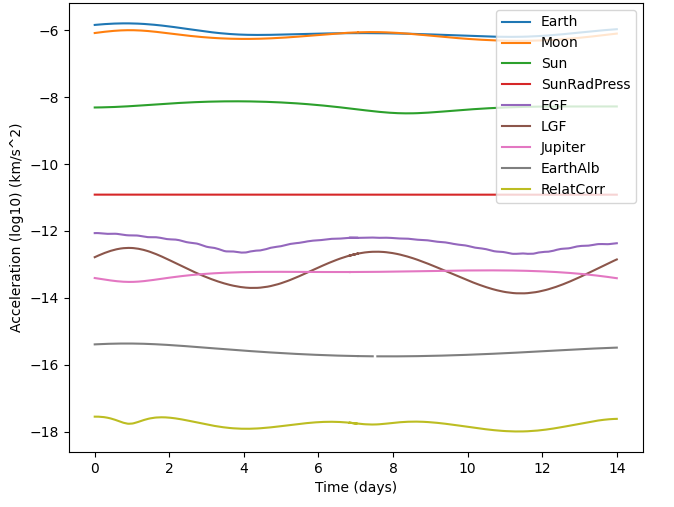}
    \caption{Magnitude of perturbations during a DRO}
    \label{fig:PartAccDRO}
\end{figure}

In Figure \ref{fig:PartAccDRO}, we study the accelerations over 1 period. Given the geometry, we should see 2 periods of the same pattern, and that is approximately what we can see for all curves. Indeed, the DRO is not exactly periodic but close to it, because we are using a high-fidelity model. The DRO is stable within a certain domain of evolution. Similar to the NRHO, the apolune and perilune are very close to the Moon's SoI, so the Earth and Moon's perturbations are of the same order of magnitude. The same applies to the associated gravitational filed correction. The relativistic correction is also very low in this situation because of the distance to the Moon.

\subsubsection{Sensitivity analysis}

In this section, we systematically remove each perturbation or improvement to evaluate the relative contribution of each. All other perturbations and improvements remain consistent with the reference scenario. As a reminder, the reference case involves a DRO with a 1-day period, using the same model parameters as the LLO described in Section \ref{RefLRO}, except for the relative tolerance, which has been adjusted to $10^{-7}$.

\begin{table}[!ht]
    \centering
    \begin{tabular}{c||c|c|c|c|c|c}
        & Reference & $10\times10$ LGF & w/o Earth & w/o Sun & w/o SRP & w/o GRC \\
        RMSE (km) & 0.418 & 0.418 & 1325 & 8.932 & 0.400 & 0.418 \\
        \hline\hline
        & $0\times0$ EGF & w/o Jupiter & w/o Albedo & ODE45 (RT=$10^{-7}$) & RT of $10^{-6}$\\
        RMSE (km) & 0.418 & 0.418 & 0.418 & 0.418 & 0.418
    \end{tabular}
    \caption{Relative role of each perturbation or improvement}
    \label{tab:relroleDRO}
\end{table}

From Table \ref{tab:relroleDRO}, we can see that removing some perturbations reduces the accuracy except for the SRP. This single anomaly may be due to an integrator's positive error or poor modeling, which will be discussed in the summary section. The results are quite similar to those of the NRHO. Overall, we can see that the biggest impact is from Earth's attraction, followed by the Sun, while the others are minimal. Further away from the Moon compared to the LLO, the integrator's parameters are less significant for accuracy, and the main influences are the two other bodies, Earth and Sun. The importance of Earth's influence is expected due to the definition of the orbit in the Earth-Moon rotational frame.

\subsection{Sum up of the perturbations over all orbits} \label{Sum Up Pert}
In the following section, we will summarize the data from the four orbits previously discussed. To enable effective comparison, a reference value will be established for each orbit, and the ratio of each perturbation's contribution to this reference value will be calculated. The reference will represent the error for a 1-day propagation, with linear adjustments applied if the orbital period exceeds one day.

\begin{table}[!ht]
    \centering
    \begin{tabular}{c||c|c|c|c}
        Orbits & LLO & ELFO & NRHO & DRO \\
        \hline \hline
        Reference 1 day (m) & 104.7 & 39.5 & 222 & 417.8 \\
        \hline
        $10\times10$ LGF    & 10.9 &  2.47 &  $\approx$ 1 & $\approx$ 1 \\
        w/o Earth     & 1.94 & 95.6 & 7143 & 3170\\
        w/o Sun       & 0.93 & 2.31 & 64.6 & 21.3\\
        w/o SRP       & 0.90 & 0.94 & 1.23 & 0.96\\
        w/o GRC       & 0.93 &  $\approx$ 1 &  $\approx$ 1 &  $\approx$ 1 \\
        $0\times0$ EGF      & 0.98 & $\approx$ 1 &  $\approx$ 1 &  $\approx$ 1 \\
        w/o Jupiter   & 1.05&  $\approx$ 1 &  1.03 &  $\approx$ 1 \\
        w/o Albedo    & 0.87 &  $\approx$ 1 &  $\approx$ 1 &  $\approx$ 1 \\
        \hline
        ODE45 (RT=$10^{-7}$)   & 1.44 &  1.32 &  0.97 & $\approx$ 1\\
        RT=$10^{-6}$ & 7.16 &  14.1 &  1.75 & $\approx$ 1
    \end{tabular}
    \caption{Sum up of all perturbations over all orbits}
    \label{tab:SU}
\end{table}

Much information can be retrieved from Table \ref{tab:SU}. The values for each perturbation help to assess the impact on the orbit. If the ratio is high, it means that the impact is significant, as removing the perturbation increases the error by a factor of the calculated ratio. If it is close to 1, the impact is small, and if it is less than 1, it is an anomaly. We can observe multiple anomalies in the table, particularly for the LLO and SRP. These anomalies suggest either a positive integration error or that the modeling needs improvement. For instance, the SRP modeling assumes a circular section for the spacecraft, which might lack accuracy. These anomalies might also be attributed to the reference orbits' accuracy and precision and the short time span of the orbits (only one day). However, we couldn't perform a wide-scale analysis due to the limitation of reference orbits, which must be from existing lunar missions. For the anomalies in the LLO orbit, we observe that the last two parameters vary significantly, corresponding to the integrator's impact. Therefore, the orbit can be considered hard to integrate, and the strong and complex LGF can explain the positive and small anomalies. However, the ELFO orbit also presents high variation for the integration-related parameters but without many anomalies. The high amplitude of the complex LGF in LLO remains the main candidate.

Moreover, in terms of the 1-day reference, the ELFO has the best accuracy, followed by the LLO, then the NRHO, and finally the DRO. For the comparison between LLO and ELFO, we can argue that the closer to the Moon, the more impact the complex lunar gravity has, which would explain why the LLO is less accurate. Then for the NRHO and DRO, we observe the significant influence of the Earth because they are TBP orbits. They are more complex, and the propagation is harder, which explains the gap with the first two orbits. However, the DRO is known to be a much more stable orbit than the NRHO, so we would have expected it to have better accuracy than the NRHO. We could explain this by the quality of the DRO reference. Indeed, it was only a 1-day sample instead of the full orbit (14 days), and there were many maneuvers and discontinuities in the trajectory. We chose a clean segment, but minor control maneuvers could still be in the chosen segment. Additionally, we observe that the LGF's influence decreases from LLO to DRO, and the influence of the Earth decreases, which is explained by the growing distance to the Moon.

Lastly, we can see that the Earth albedo, Jupiter attraction, EGF, GRC, and even SRP have consistently low impacts on all orbits except the LLO, which is more complex. This table \ref{tab:SU} can then be used to assess which forces or perturbations can be removed to be cost-efficient in computational time if the propagation has to be done over a long period.

\section{Mission design} \label{MD}
The refined propagator was then incorporated into a mission design tool, the so-called HALO. The tool provides five modes that can be arranged sequentially to form a specific mission. 

The first mode is free propagation with the propagator given a specific span. The second mode is propagation with an added maneuver: the maneuver at the initial point allows reaching the desired orbit with an impulsive change of velocity. It requires a span and a target orbit as input. This mode is particularly useful at the end of a Lambert transfer where we want to boost to the desired orbit. The third mode is the Lambert transfer between two points, using an already developed, fast, and robust solver \citep{Lambert}. It also requires a span and a target orbit. The fourth mode is an optimized Lambert solver based on the previous solver; the algorithm is presented in the next section. It requires a target orbit and a guess for the three parameters defined in the related section. Lastly, the fifth mode is a trajectory convergence algorithm for TBP orbits (DROs and NRHOs), also developed in the next sections. This last mode only needs the period of the orbit, which for a TBP is the periodic time for which the trajectory repeats in the rotational frame, as there is usually no periodicity of the trajectory in the inertial frame.

The following section will develop the algorithms used for the last two modes. The two proposed algorithms do not claim to be optimal but provide a good first solution to the problem that could be further refined if needed.

\subsection{Lambert problem optimizer algorithm}
The problem we aim to solve here is determining the optimized way to transfer from one orbit to another. This problem is particularly relevant to the study of TBP orbits because the Gateway space station is in an NRHO, and DV (Delta Velocity)-efficient transfers will need to be studied for various orbits in the cis-lunar domain to the Gateway NRHO.

To solve this problem, we use the Matlab function `fminsearch', a nonlinear programming solver that employs a simplex search method. The objective function to minimize is the sum of the DVs for the boost to enter the transfer orbit and the boost to join the second orbit, with an added penalty term. This penalty term is proportional to the norm of the difference between the Lambert target (second orbit) and the resulting propagation after the Lambert calculation. Essentially, it represents the error between the target and the final position. The penalty factor was chosen as 1 km/s for every 2 km of error. We have three parameters to optimize: the departure point from the first orbit, the transfer span, and the arrival point on the second orbit. To define a position on an orbit, we use a propagation time from a reference position on the orbit. Thus, we have three parameters: two times defining the positions on the departure and arrival orbits, and a time span for the travel time between the two points.

As an example, we optimized the transfer from the Gateway NRHO to an ELFO, as shown in Figure \ref{fig:OLambert}. The initial situation required a total DV of 4.470 km/s and had an error of 251 km, whereas the optimized situation required only 0.816 km/s of DV and had an error of 122 km.

\begin{figure}
    \centering
    \includegraphics[width=0.5\linewidth]{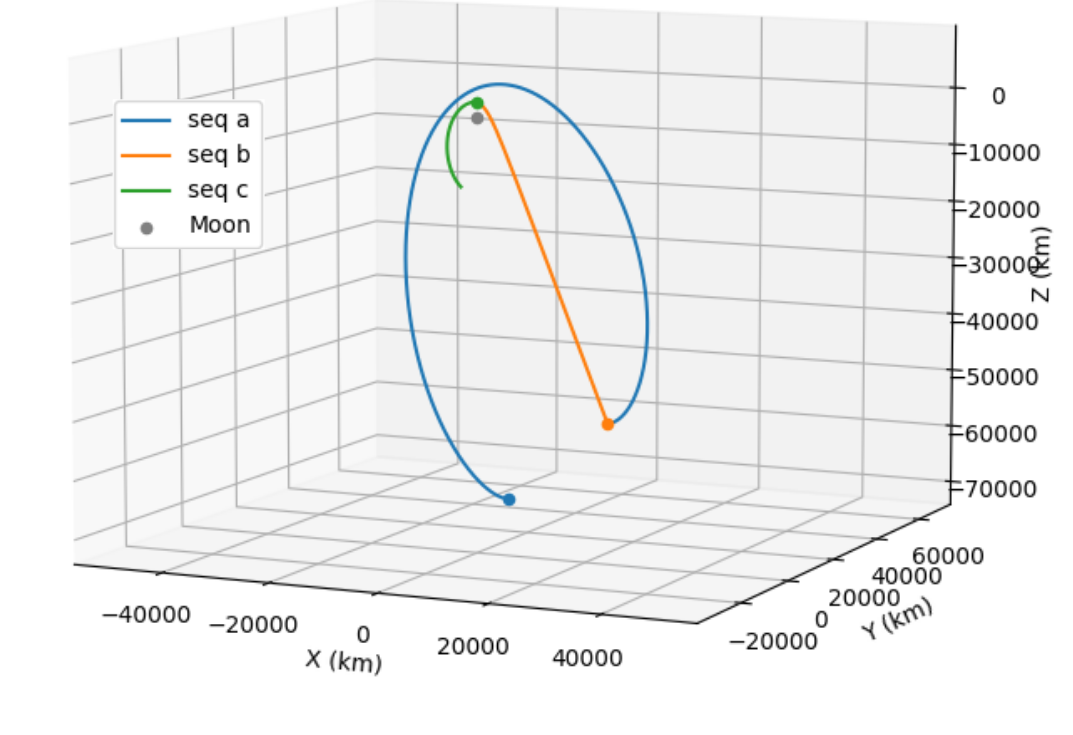}
    \caption{Initial situation of a Lambert transfer between the Gateway NRHO and an ELFO}
    \label{fig:InitL}
\end{figure}
\begin{figure}
    \centering
    \includegraphics[width=0.5\linewidth]{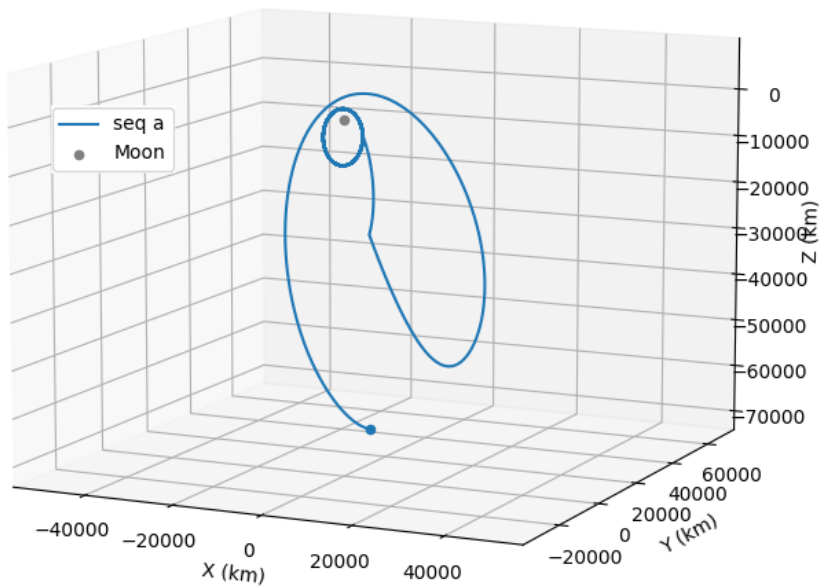}
    \caption{Converged situation of a Lambert transfer between the Gateway NRHO and an ELFO}
    \label{fig:FinalL}
\end{figure}

\subsection{Three-body problem optimization algorithm} \label{3BPOA}
When studying TBP orbits (such as DROs and NRHOs), after fitting a simplified model trajectory (calculated in the CR3BP) to an ephemeris model (as described in Section \ref{3BPO}), the trajectory is no longer in a TBP orbit. However, this fitting provides a good initial guess to converge to a high-precision model TBP orbit.

In order to improve our initial guess, we use the Matlab function `fsolve', a nonlinear system root solver that uses a Levenberg-Marquardt algorithm in our case, a nonlinear least squares algorithm. The objective function we chose to nullify is composed of multiple similar parts. The first part is the difference between the original state and the state reached after a one-period propagation, and the other parts are the same but with multiple-periods propagation. If this objective function is nullified, the same state will be reached after propagation, and the motion will be periodic. The use of the multiple-period part is to achieve better convergence because the zero is never exactly reached. We can choose the number of periods taken into account depending on the stability of the solution.

However, the TBP orbits are not periodic in the inertial frame but in the rotational one, so the conversion presented in Section \ref{3BPO} has to be done. Therefore, the decision variables are the 6 variables corresponding to the original state, and we have a system of $6n$ equations for the objective function, 6 for each periodicity after $i$ periods for $i$ from 1 to $n$, $n$ being the number of periods taken into account. We use Section \ref{3BPO} for the conversion on the first two lines, and $f$ is the objective function.

\begin{align}
    & \Vec{MS}_{[\text{R}]}(t) = M_{\text{RI}}^{-1}(t) . \Vec{MS}_{[\text{I}]}(t)\nonumber\\
    & \Vec{V}_{\text{S/M[R]}}(t) = M_{\text{RI}}^{-1}(t) .\left( \Vec{V}_{\text{S/M[I]}}(t) - \vec{\Omega}_{\text{R/I}}(t) \times \Vec{MS}_{[\text{I}]}(t) \right) \nonumber\\
    & f\left( \begin{array}{c}
        \Vec{MS}_{[\text{I}]}(t=0) \\
        \Vec{V}_{\text{S/M[I]}}(t=0)
        \end{array} \right) = \left\{
            \begin{array}{c}
                \forall i \in [\![1;n]\!] \\
                \vec{MS}_{[\text{R}]}(t=i\cdot T) - \vec{MS}_{[\text{R}]}(t=0) \\
                \Vec{V}_{\text{S/M[R]}}(t=i\cdot T) - \Vec{V}_{\text{S/M[R]}}(t=0)
            \end{array}
        \right.
\end{align}

Two examples are demonstrated: one with an NRHO similar to the Gateway orbit and another with a DRO akin to the one used by the ORION spacecraft during the Artemis I mission. First, the Gateway NRHO has a period of about 6.5 days, a perilune altitude of 1500 km, and an apolune of 70,000 km. We chose an NRHO with a similar period of 6.65 days and comparable apolune and perilune. We aim to make it converge to an accurate ephemeris trajectory on the arbitrarily chosen date of January 1, 2024 (as if we were designing a mission). After retrieving the initial data of this similar orbit in the CR3BP from the JPL database \citep{3BPO}, we can fit the initial state to the ephemeris model according to Section \ref{3BPO}, and initialize the previous optimization algorithm with it.

\begin{figure}
    \centering
    \includegraphics[width=0.5\linewidth]{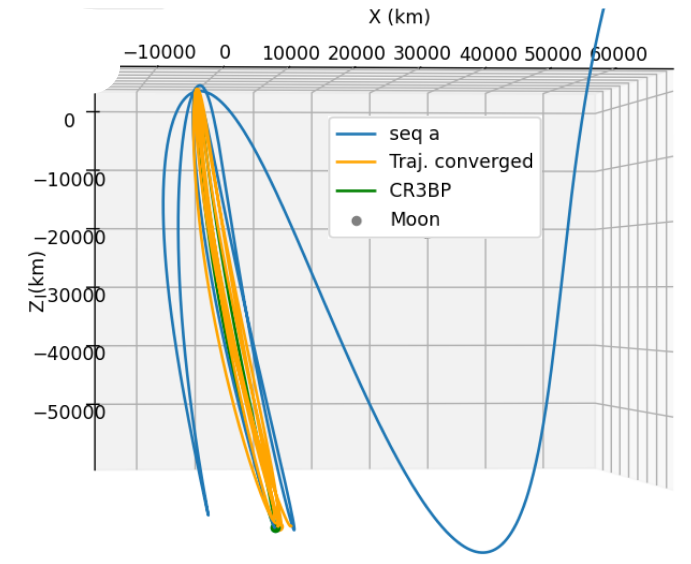}
    \caption{Optimized NRHO trajectory in the Moon centered rotational frame}
    \label{fig:NRHOC}
\end{figure}

\begin{figure}
    \centering
    \includegraphics[width=0.7\linewidth]{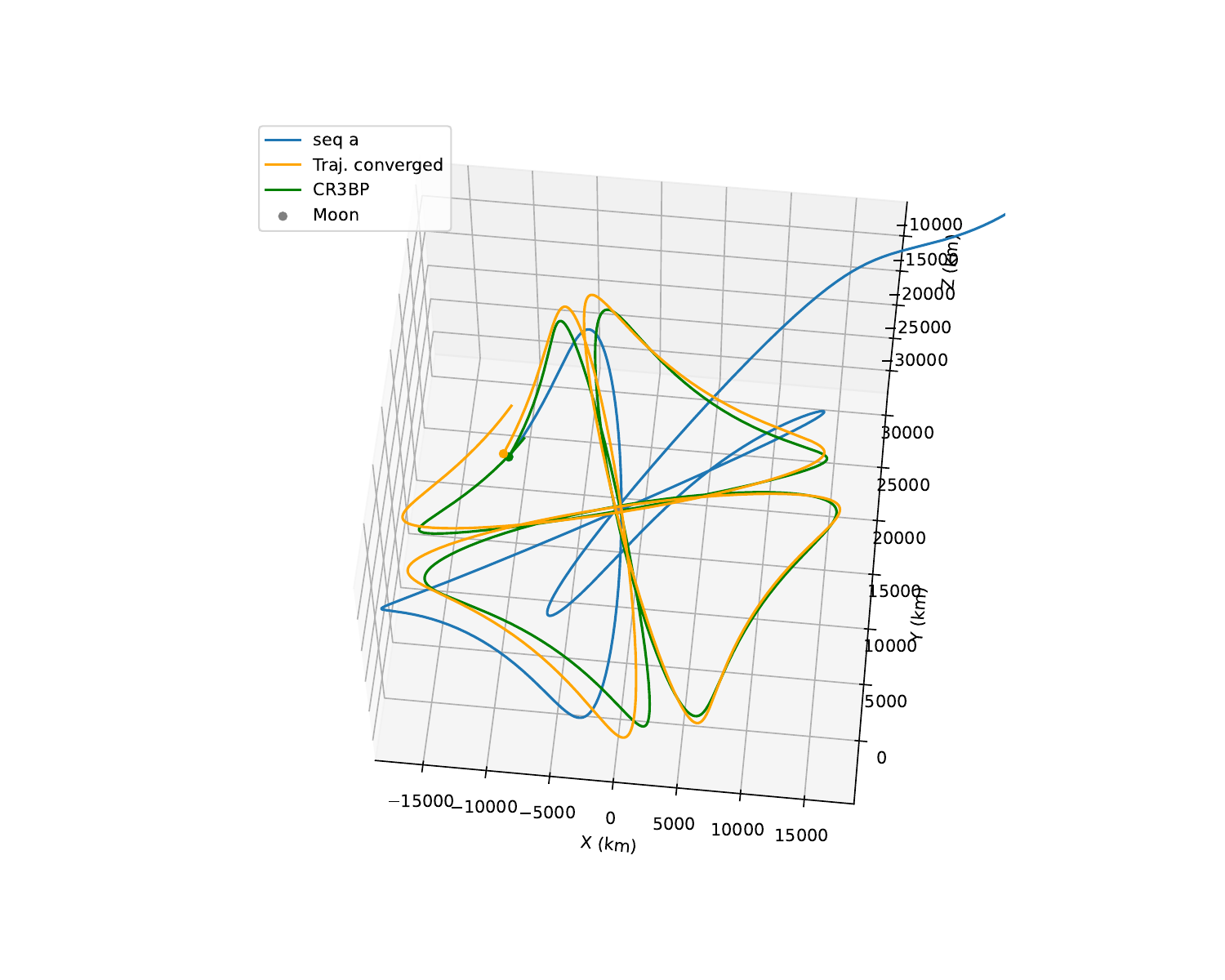}
    \caption{Optimized NRHO trajectory in the Moon centered inertial frame}
    \label{fig:NRHOCI}
\end{figure}

In Figure \ref{fig:NRHOC}, we have the CR3BP trajectory in green. The blue trajectory corresponds to the propagation of the fitted CR3BP trajectory (the initial guess), and the orange trajectory is the result of the optimization over three periods. The computation time was about 75 seconds with 31 function calls. We can see a clear improvement from the blue trajectory, which goes out of the NRHO after four periods, to the orange one, which is concentrated around the green trajectory in the rotational frame and follows the pattern of the CR3BP trajectory in the inertial frame. The orange trajectory is not exactly similar to the green one, but the change of the initial state allowed the blue trajectory to turn into an orbit (orange) with similar properties to the green one.

The second example will be with a DRO similar to the one used in the Artemis I mission. This orbit had a period of around 14 days, so we chose a CR3BP DRO with this period span from the JPL database \citep{3BPO}. Similarly, we fit the data and initialize the optimization.

\begin{figure}
    \centering
    \includegraphics[width=0.5\linewidth]{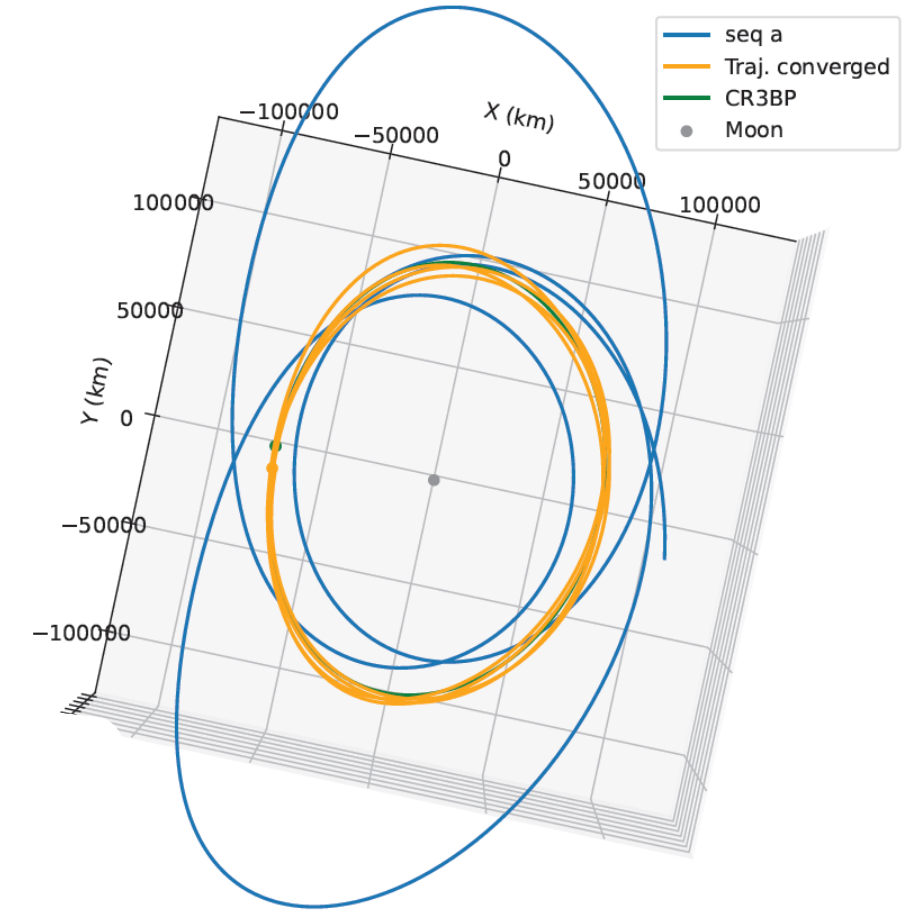}
    \caption{Optimized DRO trajectory in the Moon centered rotational frame}
    \label{fig:DROC}
\end{figure}
\begin{figure}
    \centering
    \includegraphics[width=0.7\linewidth]{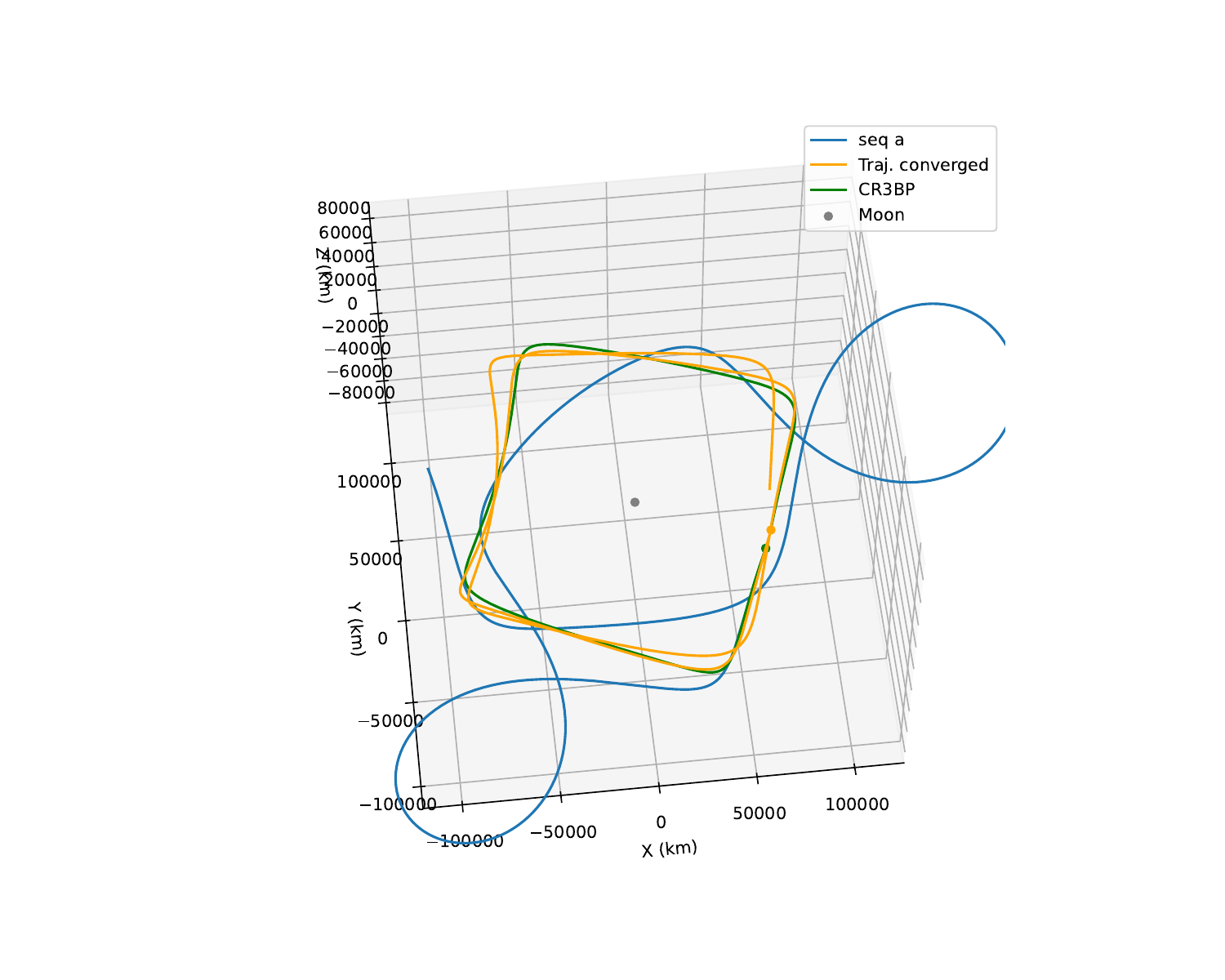}
    \caption{Optimized DRO trajectory in the Moon centered inertial frame}
    \label{fig:DROCI}
\end{figure}

For the same number of periods as the NRHO ($n=3$), the optimization does not converge well in this case. Therefore, the convergence was reached in Figures \ref{fig:DROC} and \ref{fig:DROCI} for $n=2$. This hard convergence may be due to a poor initial guess and the long period of 14 days. Thus, a convergence with $n=3$ can be processed after the first convergence if a better solution is needed.

Similar to the NRHO, the initial guess (blue) goes out of the DRO after only a few days. After half a period (so a quarter square in the inertial frame), the orange trajectory (optimized) is well closed in the rotational frame and shares the same characteristics as the CR3BP square in the inertial frame.

These two examples show how to use the algorithm and what to do if the convergence is not reached.

\section{Conclusion}
During this study, we have developed a high-accuracy propagator, provided detailed modeling of all forces and perturbations, and implemented the propagator into an open-source mission design tool, HALO. The detailed modeling and comprehensive explanation of algorithms and methods were done to allow easy understanding and quick appropriation of the tool for other studies. An assessment of the propagator on different interesting orbits was conducted in section \ref{Assess}, which provides information on the propagator and methods for users to assess it on their own.

We discussed how the modeling could be improved in the related Section \ref{Modelling} and how different modifications to the integration methods, for instance, are of prime importance for the sake of the propagator's accuracy. These are the ways to go to further improve the propagator depending on the purpose of the improvement. This purpose can be motivated by the summary of the results over different orbits in Section \ref{Assess}. Then the algorithms and use of the mission design tool were presented in Section \ref{MD}.

We chose to assess the propagator with real ephemeris data from spacecraft in the orbits of interest, but we could also have chosen to compare the accuracies with widely used mission design tools like STK or GMAT. We also chose to implement specific optimization techniques for specific mission design problems like the optimized Lambert and the optimized three-body problem trajectory.

We hope that this open-source tool will be used in other research, and we encourage further development of it.

\section{Acknowledgement}
The first author would like to acknowledge the UNSW Faculty of Engineering's funding support for my research in Australia. The first author would also like to acknowledge the work of Enio Condoleo and his Lunar High Precision Orbit Propagator which was the base work for this study.

\section{Declaration of competing interest}
The authors have no competing interests to declare that are relevant to the content of this article.

\section{References}

\bibliographystyle{plainnat}  
\bibliography{ref}

\end{document}